%% file: gnn-partitioning-study.tex
\documentclass[sigconf,edbt]{acmart-edbt2025}

\def\BibTeX{{\rm B\kern-.05em{\sc i\kern-.025em b}\kern-.08em
    T\kern-.1667em\lower.7ex\hbox{E}\kern-.125emX}}

\usepackage{booktabs} 

\newcommand{\blue}[1]{\textcolor{black}{#1}}

\usepackage{caption}
\usepackage{subcaption}
\usepackage{float}
\usepackage{multirow}
\usepackage{pifont}
\usepackage{makecell}

\newcommand{\HO}{\emph{HO}}
\newcommand{\DI}{\emph{DI}}
\newcommand{\EN}{\emph{EN}}
\newcommand{\EU}{\emph{EU}}
\newcommand{\OR}{\emph{OR}}
\newcommand{\PA}{\emph{PA}}

\newcommand{\KAHIP}{\emph{KaHIP}}
\newcommand{\SPINNER}{\emph{Spinner}}
\newcommand{\RANDOM}{\emph{Random}}
\newcommand{\METIS}{\emph{Metis}}
\newcommand{\LDG}{\emph{LDG}}
\newcommand{\BYTEGNN}{\emph{ByteGNN}}

\newcommand{\DBH}{\emph{DBH}}
\newcommand{\HDRF}{\emph{HDRF}}
\newcommand{\TWOPS}{\emph{2PS-L}}
\newcommand{\HEP}{\emph{HEP}}
\newcommand{\HEPP}{\emph{HEP10}}
\newcommand{\HEPPP}{\emph{HEP100}}

\newcommand{\DISTGNN}{\textit{\mbox{DistGNN}}}
\newcommand{\DISTDGL}{\textit{\mbox{DistDGL}}}

\setcopyright{rightsretained}

\acmDOI{}

\acmISBN{978-3-98318-097-4}

\acmConference[EDBT 2025]{28th International Conference on Extending Database Technology (EDBT)}{25th March-28th March, 2025}{Barcelona, Spain}
\acmYear{2025}

\begin{document}
\title{An Experimental Comparison of Partitioning Strategies for Distributed Graph Neural Network Training}

\author{Nikolai Merkel}
\affiliation{
  \institution{Technical University of Munich}
}
\email{nikolai.merkel@tum.de}

\author{Daniel Stoll}
\affiliation{
  \institution{Technical University of Munich}
}
\email{daniel.stoll@tum.de}

\author{Ruben Mayer}
\affiliation{%
  \institution{University of Bayreuth}
}
\email{ruben.mayer@uni-bayreuth.de}

\author{Hans-Arno Jacobsen}
\affiliation{%
  \institution{University of Toronto}
}
\email{jacobsen@eecg.toronto.edu}

\renewcommand{\shortauthors}{}

\begin{abstract}
Recently, graph neural networks (GNNs) have gained much attention as a growing area of deep learning capable of learning on graph-structured data. 
However, the computational and memory requirements for training GNNs on large-scale graphs make it necessary to distribute the training.
A prerequisite for distributed GNN training is to partition the input graph into smaller parts that are distributed among multiple machines of a compute cluster. 
Although graph partitioning has been studied with regard to graph analytics and graph databases, its effect on GNN training performance is largely unexplored.
As a consequence, it is unclear whether investing computational efforts into high-quality graph partitioning would pay off in GNN training scenarios.
  
In this paper, we study the effectiveness of graph partitioning for distributed GNN training.
Our study aims to understand how different factors such as GNN parameters, mini-batch size, graph type, features size, and scale-out factor influence the effectiveness of graph partitioning. 
We conduct experiments with two different GNN systems using vertex and edge partitioning. 
\blue{We found that high-quality graph partitioning is a very effective optimization to speed up GNN training and to reduce memory consumption.}
Furthermore, our results show that invested partitioning time can quickly be amortized by reduced GNN training time, making it a relevant optimization for most GNN scenarios.
Compared to research on distributed graph processing, our study reveals that graph partitioning plays an even more significant role in distributed GNN training, which motivates further research on the \mbox{graph partitioning problem.} 
\end{abstract}

%
%



\maketitle

\input{sections/introduction.tex}

\input{sections/background.tex}
\input{sections/experimental_setup.tex}
\input{sections/distgnn_results.tex}
\input{sections/distdgl_results.tex}

\input{sections/lessons_learnt.tex}
\input{sections/related-work.tex}
\input{sections/conclusion.tex}

\begin{acks}
This work is funded in part by the Deutsche Forschungsgemeinschaft (DFG, German Research
Foundation) - 438107855 and by the Bavarian Ministry of Economic Affairs, Regional Development and Energy (Grant: DIK0446/01).   
\end{acks}

\bibliographystyle{ACM-Reference-Format}
\bibliography{sample-base}

%
% ...

\end{document}

%% file: sections/introduction.tex
\section{Introduction}
\label{sec:introduction}
The management and processing of graph-structured data has been in the focus of research in academia and industry for many decades~\cite{pregel,powergraph,powerlyra,graphx,surveygraphprocessing}.
Graphs are an excellent way of modeling real-world phenomena that focus on the interactions and relations between entities~\cite{newman}. 
Recently, graph neural networks (GNNs) have emerged as a new category of machine learning models which are specialized for learning on graph-structured data.
GNNs have successfully been applied to domains such as recommendation systems~\cite{gnnsocial,pinsage}, natural language processing~\cite{gnnnlp}, drug discovery~\cite{gnndrug} and fraud detection~\cite{gnnfraud}.
Despite their success, GNN training is challenging due to computationally intensive deep neural network operations and high memory requirements. 
Therefore, GNNs are commonly trained in a distributed fashion. 
In order to enable distributed training, the input graph needs to be \textit{partitioned} into a predefined number of equally-sized parts which are distributed among different machines such that the cut size is minimized.

Many graph partitioning approaches exist.
In edge partitioning \cite{ne,hep,adwise,twops,dbh,hdrf,cusp}, edges are assigned to partitions while in vertex partitioning \cite{metis,spinner,ldg,bytegnn,sandersschulz2013}, vertices are assigned to partitions.
In the past decade, different works~\cite{survey.1.vldb.2017, survey.2.vldb.2018, survey.3.vldb.2018, survey.4.sigmod.2019} have studied the effect of graph partitioning on the performance of distributed graph processing systems such as Pregel~\cite{pregel}, PowerGraph~\cite{powergraph}, PowerLyra~\cite{powerlyra} or GraphX~\cite{graphx} for classical graph analytics workloads such as PageRank, Connected Components and Shortest Paths.
In these systems, a vertex function is iteratively executed for each vertex to compute state updates which are propagated along the edges to the vertices’ neighbors~\cite{pregel}. 
Often the messages and vertex states are small, e.g., consisting of a single numerical value and light-weight vertex functions such as computing the maximum value or sum of values.
Compared to GNN training, these workloads are light-weight in terms of computation and memory overhead, and are executed on the graph structure only. 
GNN training has unique features such as heavy-weight neural network operations, intensive communication of high dimensional feature vectors and large intermediate embeddings, and a large memory footprint to store intermediate results for each layer during forward and backward propagation. 
Further, GNNs have additional parameters such as number of layers, number of hidden dimensions or mini-batch size which influence memory, \mbox{computation and communication overhead.} 

The effectiveness of graph partitioning has not been investigated for distributed GNN training.
As a consequence, it is unclear under which conditions it is beneficial to invest computational resources to yield a high-quality graph partitioning.
In this paper, we perform an extensive experimental analysis of graph partitioning for distributed GNN training to close this gap and make the following contributions:

\textbf{(1)} Based on extensive evaluations with two prominent state-of-the-art distributed GNN systems (\textit{DistGNN} \cite{DistGNN} and \textit{DistDGL} \cite{distdgl}), 12 graph partitioning algorithms (edge partitioning and vertex partitioning), different GNN model architectures, different GNN parameters, and graphs of various categories with different feature sizes, \blue{we find that high-quality graph partitioning is effective for GNN training. 
Compared to random partitioning, e.g., the \HEPPP{} partitioner~\cite{hep} leads to speedups of up to 10.41x (on average 4.36x) on \DISTGNN{} and the \KAHIP{} partitioner~\cite{sandersschulz2013} leads to speedups of up to 3.47x (on average 1.37x) on \DISTDGL{}. 
Further, the memory footprint can be decreased by up to 85.1\% and on average by 51.4\% if \HEPPP{} is used for partitioning} which is crucial for memory-bound GNN training. 
This shows the enormous potential of graph partitioning for GNN workloads. 

\textbf{(2)} We show that partitioning quality properties such as the replication factor or vertex balance influence the GNN training a lot. 
We find a strong correlation between replication factor, network communication and memory footprint. 
Therefore, minimizing the replication factor is crucial for efficient distributed GNN training.
We show that vertex imbalance can decrease the speedup and can also lead to severe imbalances regarding memory utilization. 
This motivates the need for further research on balanced graph partitioning algorithms.

\textbf{(3)} We find that GNN parameters such as the hidden dimension, the number of layers, the mini-batch size and the feature size influence the effectiveness of graph partitioning, both in terms of training time and memory overheads. 
Our experiments further show that a higher scale-out factor can decrease the effectiveness of vertex partitioning, while the effectiveness increases for edge partitioning. 
These insights help to understand the interplay between GNN parameters and partitioner effectiveness.  

\textbf{(4)} We find that invested partitioning time can be amortized by faster GNN training in typical scenarios, making graph partitioning relevant for production systems. 
    This is different from classical graph processing, where graph partitioning can often not be amortized and fast streaming-based graph partitioning \mbox{performs best~\cite{ease}.} \\

Our paper is organized as follows. 
In Section~\ref{sec:background}, we introduce graph partitioning and graph neural networks. 
In Section~\ref{sec:methodology}, we describe our methodology. 
Then, we analyze the results for \mbox{DistGNN} in Section \ref{sec:results:distgnn} and for \mbox{DistDGL} in Section~\ref{sec:results:distdgl}. In Section~\ref{sec:lessons}, we summarize our main findings and in Section~\ref{sec:related-work} we discuss related work. Finally, we conclude our paper in Section~\ref{sec:conclusion}.

%% file: sections/background.tex
\section{Background}
\label{sec:background}
Let $G = (V, E)$ be a graph consisting of a set of vertices $V$ and a set of edges $E \subseteq V \times V$. 
$N(v)$ represents the set of vertices that are connected to $v$. 
In the following, we discuss graph partitioning in Section~\ref{sec:background:partitioning} and distributed GNN training in Section~\ref{sec:background:gnns}.
\subsection{Graph Partitioning}
\label{sec:background:partitioning}
\begin{figure}[t]
\centering
\includegraphics[width=0.9\columnwidth]{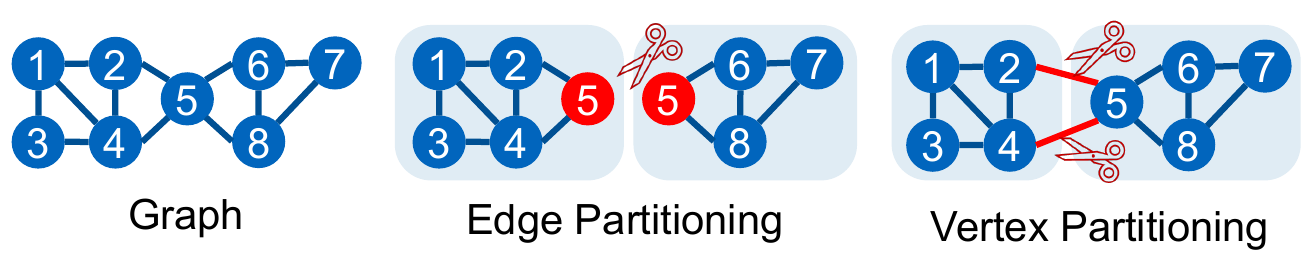}
\caption{Edge partitioning vs. vertex partitioning.}    
\label{fig:back:vertex-edge-partitionoing}
\end{figure}

The main approaches for graph partitioning are \textit{edge partitioning} and \textit{vertex partitioning} (see Figure~\ref{fig:back:vertex-edge-partitionoing}). 
In the following, we present both approaches and commonly used \mbox{partitioning quality metrics.} 

\noindent\textbf{Edge Partitioning.}
In edge partitioning (vertex-cut), the set of edges $E$ is divided into $k$ partitions by assigning each edge to exactly one partition $p \in P = \{p_1, \dots, p_k\}$ with $\cup_{i=1}^{k}p_{i} = E$. 
Through this process, vertices can be cut. 
A cut vertex is replicated to all partitions that have adjacent edges. 
\blue{Each partition $p_i$ covers a set of vertices $V(p_i) = \{ v \in V \:|  (\exists e=(u,v) \in p_i) \vee (\exists e=(v,u) \in p_i) \}$.}
The goal of edge partitioning~\cite{ne} is to minimize the number of cut vertices while keeping the partitions' edges $\alpha$-balanced, meaning $\forall p_i \in P: |p_i| \leq \alpha \cdot \frac{|E|}{k}$. 

Commonly used quality metrics to evaluate edge partitioners are the mean \textit{replication factor} and \textit{edge balance}. 
The replication factor $\mathit{RF}(P)$ is defined as $\frac{1}{|V|}\sum_{i=1}^k|V(p_i)|$ and represents the average number of partitions to which vertices are replicated. 
This metric is closely related to communication cost because replicated vertices need to synchronize their state via the network. 
Additionally, it indicates memory overhead, as for each replicated vertex, the state is replicated. 
This becomes particularly significant in the context of GNN training where vertex states are substantial in size, a concern less emphasized in classical graph processing, where memory constraints are not as pivotal.
The edge balance is defined as $\mathit{EB}(P)=\frac{\mathit{max}({\{|p_1|, \dots, |p_k|\}})}{\mathit{mean}({\{|p_1|, \dots, |p_k|\}})}$, and the vertex balance as $\mathit{VB}(P)=\frac{\mathit{max}({\{|V(p_i)|, \dots, |V(p_i)|\}})}{\mathit{mean}({\{|V(p_i)|, \dots, |V(p_i)|\}})}$. 
Most edge partitioners do not explicitly balance vertices because the computational load of many graph algorithms is believed to be proportional to the number of edges as messages are aggregated along the edges (e.g., PageRank).

\noindent\textbf{Vertex Partitioning.}
In vertex partitioning (edge-cut), the set of vertices $V$ is divided into $k$ partitions by assigning each vertex $v$ to exactly one partition $p \in P = \{p_1, \dots, p_k\}$ with $\cup_{i=1}^{k}p_{i} = V$. 
\blue{An edge $e=(u,v)$ is cut if both $u$ and $v$ are assigned to different partitions.}
Vertex partitioning aims to minimize the number of cut edges while balancing the partition sizes in terms of number of vertices. 
We define $E_{cut}$ as the set of cut edges. 

Commonly used vertex partitioning quality metrics to evaluate vertex partitioners are the \textit{edge-cut ratio} and \textit{vertex balance}.
The edge-cut ratio is defined as $\lambda=\frac{|E_{cut}|}{|E|}$ and indicates communication costs, as messages are sent via edges, and cut edges lead to network communication between machines.
The vertex balance is defined as $\mathit{VB}(P)=\frac{\mathit{max}({\{|p_1|, \dots, |p_k|\}})}{\mathit{mean}({\{|p_1|, \dots, |p_k|\}})}$ and indicates computation balance. 

\noindent\textbf{Partitioner Types.}
Both edge and vertex partitioning algorithms can be categorized into 
(1)~\textit{streaming partitioners}, which stream the graph and directly assign vertices or edges to partitions \cite{dbh,hdrf,twops,adwise,ldg}. 
Streaming partitioners can be further divided into \textit{stateless} partitioners, which do not keep any state, and \textit{stateful} streaming partitioners, which use some state, e.g., the current load per partition or to which partition vertices or edges were assigned. The state is considered for the assignments. 
(2)~\textit{In-memory partitioners} which load the complete graph into the memory \cite{metis,spinner,ne,dne,sandersschulz2013}. 
For edge partitioning, there exists also a \textit{hybrid partitioner} that partitions one part with an in-memory partitioner and the remaining part with a streaming partitioner~\cite{hep}. 

\subsection{Graph Neural Network Training}
\label{sec:background:gnns}
Graph neural networks (GNNs) are a class of neural networks which operate on graph-structured data. 
GNNs iteratively learn on graphs by aggregating the local neighborhoods of vertices.
At start, each vertex $v$ is represented by its feature vector $h^{(0)}_v$.
In each layer, a vertex $v$ aggregates learned representations of its neighbors $N(v)$ of the previous layers, resulting in $a_v^{(k)} = \mathit{AGGREGATE}^{(k)}(\{h_u^{(k-1)} | u \in N(v)\})$.
Based on $a_v^{(k)}$ and its previous intermediate representation $h_v^{(k-1)}$ in layer $k-1$, vertex $v$ updates its representation to $h_v^{(k)} = \mathit{UPDATE}^{(k)}(a_v^{(k)},h_v^{(k-1)})$.  
 

The main approaches to train GNNs are \textit{full-batch} and \textit{mini-batch} training \cite{besta2022parallel}. 
In full-batch training, the entire graph is used to update the model once per epoch. In mini-batch training, each epoch contains multiple iterations, where a  mini-batch is sampled from the graph and used for training and model update.

%% file: sections/experimental_setup.tex
\section{Experimental Methodology}
\label{sec:methodology}
Our research aims to investigate the effect of graph partitioning on the performance of distributed GNN training. 
We want to answer the following five research questions: 
\begin{description}
    \item[RQ-1] How effective is graph partitioning for distributed GNN training in reducing training time and memory footprint?
    \item[RQ-2] Do classical partitioning quality metrics accurately describe the effectiveness of partitioning algorithms for distributed GNN workloads? Which partitioning quality metric is most crucial? 
    \item[RQ-3] How much is the partitioning effectiveness influenced by GNN parameters such as the \textit{number of layers}, \textit{hidden dimension}, \textit{feature size}, \textit{mini-batch size}, and \textit{type of graph}? 
    \item[RQ-4] What is the impact of the scale-out factor (number of machines used for training) on the partitioning effectiveness?
    \item[RQ-5] Can the invested partitioning time be amortized by a reduced GNN training time?
\end{description}

\blue{To address these research questions, we conduct various experiments with two distributed GNN systems \textit{DistGNN}~\cite{DistGNN} and \textit{DistDGL}~\cite{distdgl}. 
Both systems are extensions of the \textit{Deep Graph Library~(DGL)}, which is a leading framework for training GNNs on single machines.
Consequently, \DISTDGL{} and \DISTGNN{} are based on a shared codebase. 
Both systems add functionality for distributed training with  \DISTDGL{} focusing on mini-batch training and \DISTGNN{} on full-batch training. 
A key distinction between the two systems is their support for different graph partitioning methods: \DISTGNN{} utilizes edge partitioning (vertex-cut), whereas \DISTDGL{} employs vertex partitioning (edge-cut). 
In contrast to \DISTGNN{}, \DISTDGL{} is much more mature, well-documented, highly optimized, still under active development, and also used in industry.}

In the following, we introduce the datasets and metrics we use to evaluate both systems.

\noindent\textbf{Datasets.}
We selected five graphs (see Table~\ref{tab:datasets}) from the following different categories (1)~web, (2)~social, (3)~collaboration, (4)~road, and (5)~wiki to investigate the influence of the graph type on the partitioning effectiveness.  
Compared to large-scale distributed graph processing, all datasets only have a medium-sized graph structure. 
However, GNN workloads are much more communication, memory, and compute-intensive because the vertices have large feature vectors, and the computation includes heavy-weight neural network computations with large intermediate states.
Hence, graphs of this size already require heavy-weight processing for GNN training. 
In order to systematically investigate the influence of the feature size on the partitioning effectiveness, we set the feature size to commonly used values between 16 and 512.

\begin{table}[t]
\caption{Graphs of different types. \textit{Dir.} indicates if the graph is directed or undirected.}
\label{tab:datasets}
\begin{center}
\begin{tabular}{|l|l|l|l|l|}
\hline
\textbf{Graph} & \textbf{Type} & \textbf{Dir.} & {$\mathbf{|E|}$} & $\mathbf{|V|}$\\
\hline
\hline
Hollywood-2011 (\HO{}) \cite{BoVWFI,BRSLLP} & Colla. & no &229M &2M\\
\hline
Dimacs9-USA (\DI{}) \cite{konect} &Road&yes&58M &24M \\
\hline
Enwiki-2021 (\EN{}) \cite{BoVWFI,BRSLLP}&Wiki&yes&150M& 6M \\
\hline
Eu-2015-tpd (\EU{}) \cite{BoVWFI,BRSLLP,BMSB} &Web&yes&166M& 7M\\
\hline
Orkut (\OR{}) \cite{snapnets} &Social&no& 234M& 3M\\
\hline
\end{tabular}
\end{center}
\end{table}

\begin{table}[t]
\caption{Different GNN hyper-parameters for both systems.}
\label{tab:hyper-parameters}
\begin{center}
\begin{tabular}{|l||l|l|l|l|}
\hline
\textbf{Hyper-parameter} & \textbf{Values}\\
\hline
\hline
Hidden Dimension & 16, 64, 512\\
\hline
Feature size & 16, 64, 512 \\
\hline
Number of layers & 2, 3, 4\\
\hline
\end{tabular}
\end{center}
\end{table}

\noindent\textbf{Infrastructure and infrastructure metrics.}
We use a cluster composed of 32 machines.
Each machine is equipped with 64GB memory and 8 CPU cores (Intel Core Haswell, 2.4 GHz).
We measure the CPU and memory utilization and network traffic of each worker. 

\noindent\textbf{Performance metrics.}
We compare the effectiveness of graph partitioners in reducing GNN training time based on the achieved speedup over random partitioning. 
The speedup of a graph partitioner $\mathit{p}$ is defined as $\mathit{speedup(p)}=\frac{T_{random}}{T_{\mathit{p}}}$ with $T_{p}$ and ${T_{\mathit{random}}}$ being the GNN training time of partitioner $\mathit{p}$ and random partitioning, respectively.

%% file: sections/distgnn_results.tex
\section{DistGNN - Edge Partitioning}
\label{sec:results:distgnn}
\subsection{Experiments}
\label{sec:distgnn:experiments}
\textbf{Graph partitioning algorithms.} 
We use six state-of-the-art edge partitioners from different categories:
(1)~stateless streaming partitioning with \textit{random partitioning} and \DBH{} \cite{dbh},  
(2)~stateful streaming with \HDRF{} \cite{hdrf} and \TWOPS{} \cite{twops}, and 
(3)~hybrid-partitioning with \HEP{} \cite{hep}. 
\HEP{} uses a parameter $\tau$ which influences how much of the graph is partitioned in a streaming fashion and how much in memory. 
A larger value for $\tau$ leads to a better partitioning quality as a larger part of the graph is partitioned in memory. 
As suggested by the authors, we set $\tau$ to 10 and to 100, and we treat these two configurations as two different partitioners \HEPP{} and \HEPPP{} in our evaluations. \HEPPP{} corresponds to in-memory partitioning, as the part not loaded into the memory is negligible if $\tau$ is set to $100$. 

\noindent\textbf{Workloads.} 
DistGNN currently only supports GraphSage \cite{hamilton2017inductive}, one of the most common model architectures \cite{distdgl, DistGNN, deepgalois, su2021adaptive}. 
We reviewed the GNN literature to identify commonly used hyper-parameters and chose the following to cover the ranges accordingly. 
We vary the \textit{hidden dimension} from 16 to 512, the \textit{number of layers} from 2 to 4, and the \textit{feature size} from \mbox{16 to 512 (see Table~\ref{tab:hyper-parameters}).}

\noindent\textbf{Partitioning metrics.}
We measure the well-known edge partitioning quality metrics \textit{replication factor},  \textit{edge balance} and \textit{vertex balance} (see Section~\ref{sec:background:partitioning}). 

\noindent\textbf{Training \& partitioning time.}
We measure the time per epoch to compute the speedup over random partitioning (see. Section~\ref{sec:methodology}).
Furthermore, we measure the partitioning time.

\subsection{Partitioning Performance}
\label{distgnn:part-metrics}
We compare the partitioning algorithms regarding their communication costs and computational balance in the following.

\begin{figure}[t]
\centering
\begin{subfigure}[b]{0.99\linewidth}
\centering
\includegraphics[width=\linewidth]{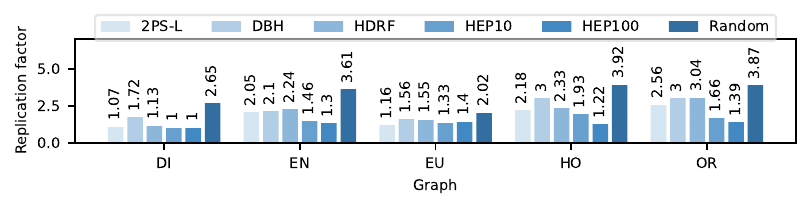}
\caption{4 partitions.}
\label{fig:distgnn:replicationfactor:4}
\end{subfigure}
\centering
\begin{subfigure}[b]{0.99\linewidth}
\centering
\includegraphics[width=\linewidth]{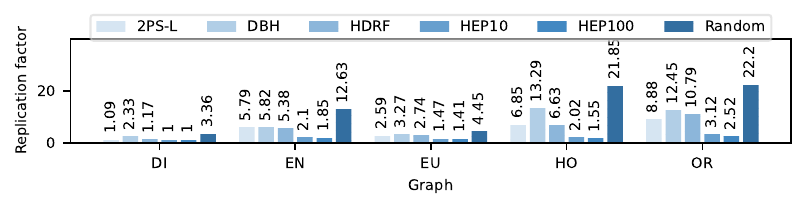}
\caption{32 partitions.}
\label{fig:distgnn:replicationfactor:32}
\end{subfigure}
\caption{Replication factors. }
\label{fig:distgnn:replicationfactor}
\end{figure}

\begin{figure}[t]
\centering
\begin{subfigure}[b]{0.32\linewidth}
\centering
\includegraphics[width=\linewidth]{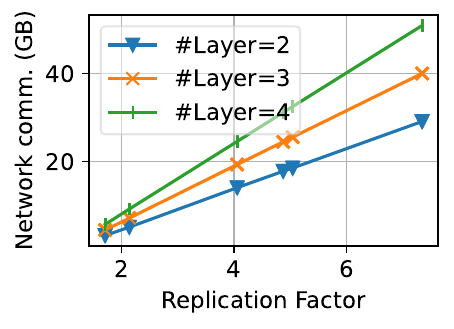}
\caption{8 machines.}
\label{fig:distgnn:repl:vs:network:8}
\end{subfigure}
\hfill
\begin{subfigure}[b]{0.32\linewidth}
\centering
\includegraphics[width=\linewidth]{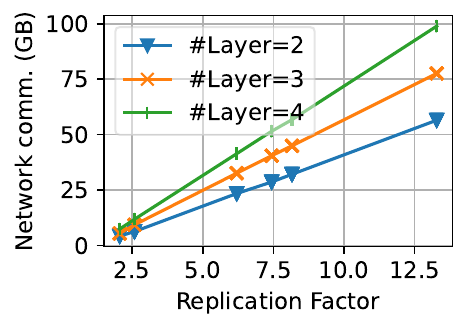}
\caption{16 machines.}
\label{fig:distgnn:repl:vs:network:16}
\end{subfigure}
\hfill
\begin{subfigure}[b]{0.32\linewidth}
\centering
\includegraphics[width=\linewidth]{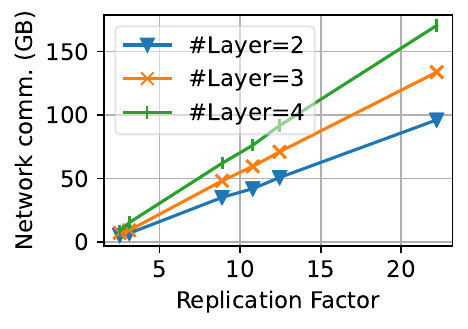}
\caption{32 machines.}
\label{fig:distgnn:repl:vs:network:32}
\end{subfigure}
\caption{Replication factor vs. network communication for different number of machines and number of layers on \OR{}.}
\label{fig:distgnn:repl:vs:network}
\end{figure}

\textit{(1) Communication costs.} 
We observe significant differences in terms of replication factors for the different partitioners. 
In all cases, \HEPPP{} leads to the lowest (best) replication factor and \RANDOM{} to the largest (worst) one. 
Figure~\ref{fig:distgnn:replicationfactor:32} shows, for example, that \HEPPP{} at 32 partitions leads to a replication factor of $2.52$ on \OR{}, which is much smaller compared to \RANDOM{} which leads to a replication factor of $22.2$. 
In general, more partitions lead to larger replication factors.
For some partitioners, the replication factors increase more sharply than for others if the number of partitions increases. 
We observe a strong correlation ($R^2 \geq 0.98$) between the replication factor and network traffic. This correlation is shown in Figure~\ref{fig:distgnn:repl:vs:network} for \OR{} for different numbers of machines and number of layers and is also observed for the remaining graphs. 
The observation is plausible. 
The higher the replication factor, the more data is communicated via the network because more vertices are replicated and must synchronize their states. 
We conclude that \textbf{minimizing the replication factor is crucial for reducing network overhead}. 

\begin{figure}[t]
\centering
\begin{subfigure}[b]{0.99\linewidth}
\centering
\includegraphics[width=\linewidth]{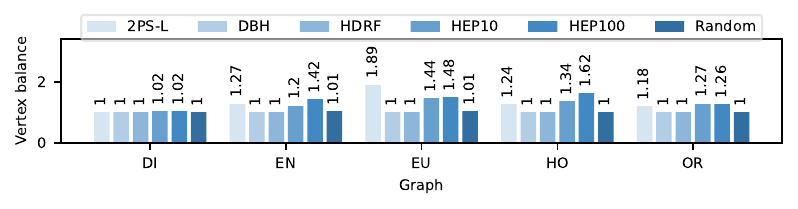}
\caption{4 partitions.}
\label{fig:distgnn:vertex-balance:4}
\end{subfigure}
\centering
\begin{subfigure}[b]{0.99\linewidth}
\centering
\includegraphics[width=\linewidth]{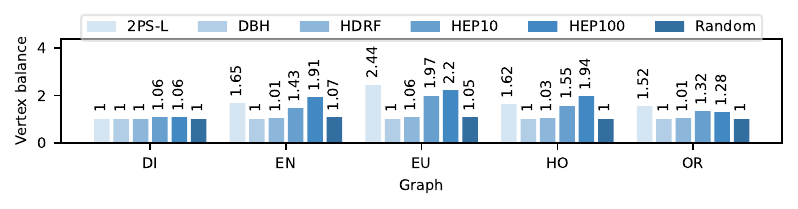}
\caption{32 partitions.}
\label{fig:distgnn:vertex-balance:32}
\end{subfigure}
\caption{Vertex balance.}
\label{fig:distgnn:vertex-balance}
\end{figure}

\begin{figure}[t]
\centering
\includegraphics[width=\columnwidth]{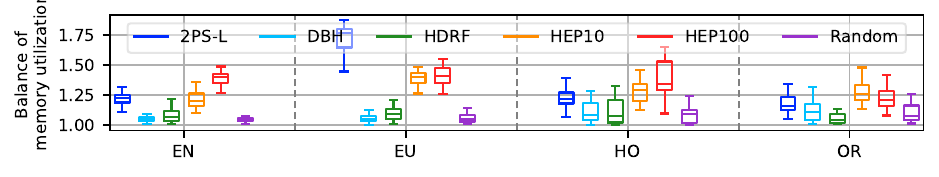}
\caption{Memory utilization balance (4 machines).}    
\label{fig:distgnn:memory-imbalance}
\end{figure}

\textit{(2) Computational balance.}
It is crucial to balance the number of edges and vertices per partition. 
Each edge leads to an aggregation in the GNN, and neural network operations are performed for vertices. 
We observe a good edge balance of at most $\alpha \leq 1.11$ for all partitioners.
However, we observe significant vertex imbalances (see Figure~\ref{fig:distgnn:vertex-balance}). 
Especially, the partitioners \TWOPS{}, \HEPP{} and \HEPPP{} lead to large vertex imbalances between $1.18$ and $1.89$ on 4 machines (see. Figure~\ref{fig:distgnn:vertex-balance:4}) and can even increase up to 2.44 on 32 machines (see. Figure~\ref{fig:distgnn:vertex-balance:32}).
The vertex imbalance has a significant influence on the balance of memory utilization.
Figure~\ref{fig:distgnn:memory-imbalance} reports the imbalance regarding memory utilization for all combinations of \textit{feature size}, \textit{hidden dimension}, and \textit{number of layers} per graph and graph partitioner.
We observe that vertex imbalance correlates with memory utilization imbalance.
We conclude that \textbf{minimizing vertex imbalance is crucial for balancing memory utilization}.

\begin{figure}[t]
\centering
\begin{subfigure}[b]{0.99\linewidth}
\centering
\includegraphics[width=\linewidth]{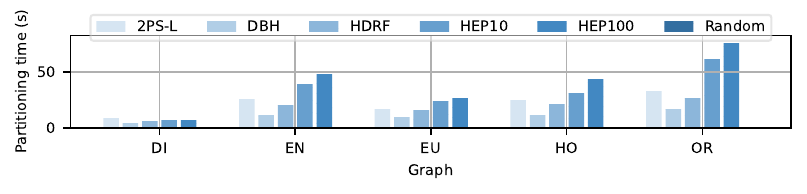}
\caption{4 partitions.}
\label{fig:distgnn:partitioningtime:4}
\end{subfigure}
\centering
\begin{subfigure}[b]{0.99\linewidth}
\centering
\includegraphics[width=\linewidth]{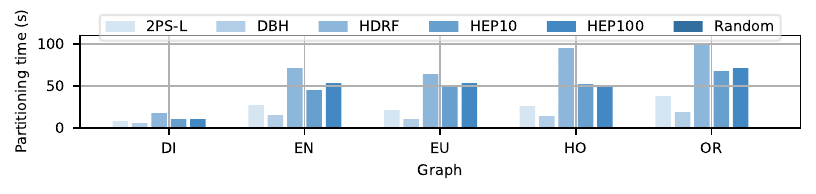}
\caption{32 partitions.}
\label{fig:distgnn:partitioningtime:32}
\end{subfigure}
\caption{Partitioning time.}
\label{fig:distgnn:partitioningtime}
\end{figure}

\textit{(3) Partitioning time.}
Figure~\ref{fig:distgnn:partitioningtime} reports the partitioning time for partitioning all graphs into 4 and 32 partitions. 
We observe that the partitioning times of some algorithms, e.g., \RANDOM{}, \TWOPS{} and \DBH{}, are less dependent on the number of partitions, compared to the remaining partitioners, where more partitions lead to higher partitioning times. 
For example, \HDRF{} takes much more time to partition the graphs into 32 partitions compared to 4 partitions, which is expected because the complexity of the scoring function depends on the number of partitions.

\subsection{GNN Training Performance}
\label{sec:distgnn:influence-gnn-parameters}
\begin{figure}[t]
\centering
\begin{subfigure}[b]{0.99\linewidth}
\centering
\includegraphics[width=\linewidth]{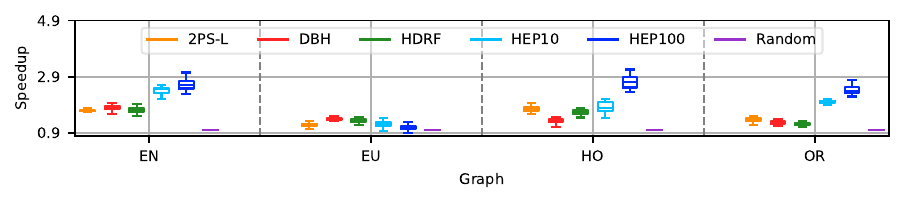}
\caption{4 machines.}
\label{fig:distgnn:overview-epoch-times:4}
\end{subfigure}
\centering
\begin{subfigure}[b]{0.99\linewidth}
\centering
\includegraphics[width=\linewidth]{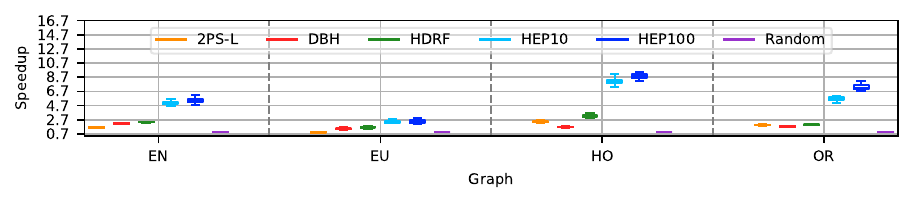}
\caption{32 machines.}
\label{fig:distgnn:overview-epoch-times:32}
\end{subfigure}
\caption{Speedup distribution of graph partitioners on 4 and 32 machines for all experiments. }
\label{fig:distgnn:overview-epoch-times}
\end{figure}

In Figure~\ref{fig:distgnn:overview-epoch-times}, we report the speedup distribution for all partitioners compared to random partitioning for all combinations of \textit{feature size}, \textit{hidden dimension}, and \textit{number of layers} (see Table~\ref{tab:hyper-parameters}). 

\begin{figure}[t]
\centering
\begin{subfigure}[b]{0.32\linewidth}
\centering
\includegraphics[width=\linewidth]{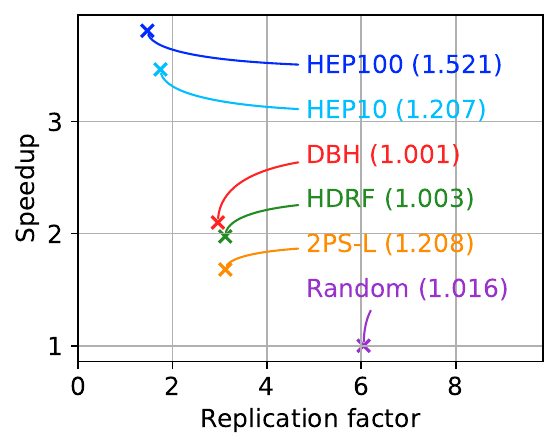}
\caption{8 machines.}
\label{fig:distgnn:repl:vs:speedup:8}
\end{subfigure}
\hfill
\begin{subfigure}[b]{0.32\linewidth}
\centering
\includegraphics[width=\linewidth]{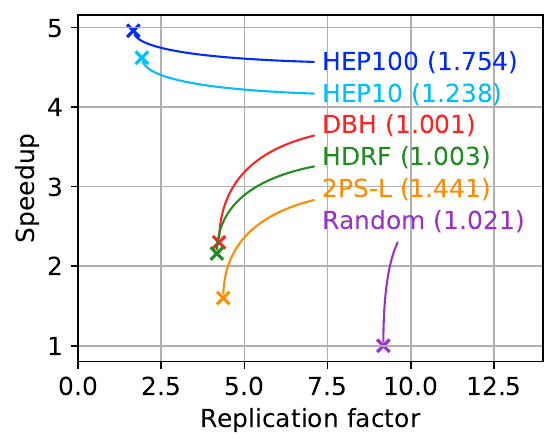}
\caption{16 machines.}
\label{fig:distgnn:repl:vs:speedup:16}
\end{subfigure}
\hfill 
\begin{subfigure}[b]{0.32\linewidth}
\centering
\includegraphics[width=\linewidth]{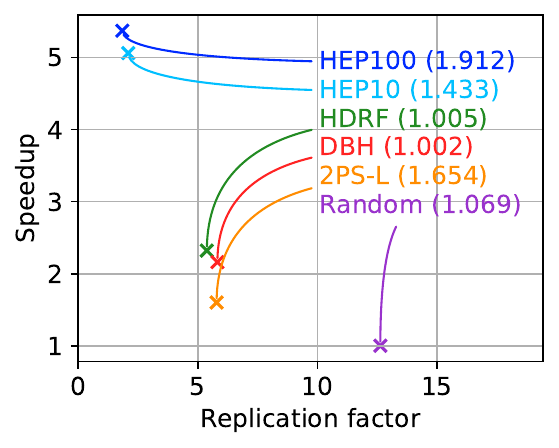}
\caption{32 machines.}
\label{fig:distgnn:repl:vs:speedup:32}
\end{subfigure}
\caption{Replication factor vs. speedup on \EN{} along with the vertex balance in the brackets. }
\label{fig:distgnn:repl:vs:speedup}
\end{figure}

\begin{figure}[t]
\centering
\begin{subfigure}[b]{0.24\linewidth}
\centering
\includegraphics[width=\linewidth]{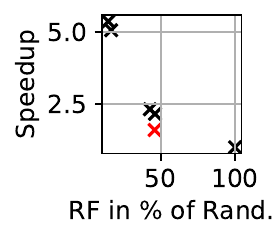}
\caption{EN.}
\label{fig:distgnn:repl:vs:speedup:norm:EN}
\end{subfigure}
\hfill
\begin{subfigure}[b]{0.24\linewidth}
\centering
\includegraphics[width=\linewidth]{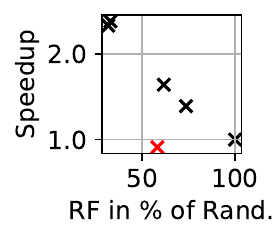}
\caption{EU.}
\label{fig:distgnn:repl:vs:speedup:norm:EU}
\end{subfigure}
\hfill 
\begin{subfigure}[b]{0.24\linewidth}
\centering
\includegraphics[width=\linewidth]{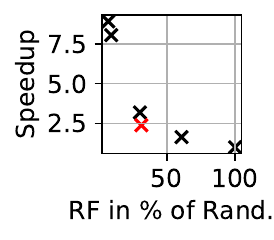}
\caption{HO.} 
\label{fig:distgnn:repl:vs:speedup:norm:HO}
\end{subfigure} 
\hfill 
\begin{subfigure}[b]{0.24\linewidth}
\centering
\includegraphics[width=\linewidth]{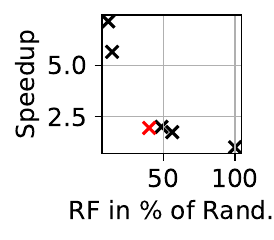}
\caption{OR.}
\label{fig:distgnn:repl:vs:speedup:norm:OR}
\end{subfigure}
\caption{Replication factor (RF) in \% of random partitioning (Rand.) vs. speedup on 32 machines. \TWOPS{} is highlighted red.}
\label{fig:distgnn:repl:vs:speedup:norm}
\end{figure}

We observe that graph partitioning is effective in reducing the training time. 
\HEPPP{} leads to the largest speedups of up to $3.53x$, $6.18x$, $8.15x$ and $10.41x$ on the graphs \EU, \EN{}, \OR{} and \HO, respectively.
However, the effectiveness of the partitioners differs. 
In most cases, the more lightweight streaming-based partitioners \DBH, \TWOPS{} and \HDRF{} yield smaller speedups compared to more heavy-weight partitioning with \HEPP{} and \HEPPP.
For example, on \OR, compared to \RANDOM, on average \DBH, \TWOPS, \HDRF, \HEPP{} and \HEPPP{} lead to speedups of $1.40x$, $1.46x$, $1.44x$, $2.96x$ and $3.68x$ on 8 machines, $1.62x$, $1.61x$, $1.75x$, $4.37x$ and $7.16x$ on 16 machines and $1.74x$, $1.95x$, $2.00x$, $5.67x$ and $7.16x$ on 32 machines. 
Further, we observe that the effectiveness of the partitioners is influenced by the graph type, e.g., the speedups for the web graph \EU{} which has a lower density than the remaining graphs, are smaller for all partitioners, compared to the remaining graph types. 
We also observe for \EU{}, that the partitioners are less effective in reducing the replication factor over random partitioning and that the replication factors are lower - even under random partitioning - compared to the other graphs.
In the Figures~\ref{fig:distgnn:repl:vs:speedup:norm:EN}-\ref{fig:distgnn:repl:vs:speedup:norm:OR}, we plot the replication factor in percent of random partitioning against the speedup for \EN{}, \EU{}, \HO{} and \OR{}, respectively. 
Each marker represents a graph partitioner, and \TWOPS{} is highlighted in red.
A trend emerges from the data: a lower replication factor relative to the replication factor of random partitioning correlates with a higher speedup. 
Consequently, it's evident that to achieve large speedups, it's vital to minimize the replication factor.
Nonetheless, exceptions such as \TWOPS{} on \EU{} (see Figure~\ref{fig:distgnn:repl:vs:speedup:norm:EU}) do exist.  
Such deviations are often due to a significant vertex imbalance.
The vertex imbalance leads to slowdowns for \TWOPS{} of 0.92x, 0.92x, and 0.91x on 8, 16 and 32 machines on \EU.  
Another example can be seen in Figure~\ref{fig:distgnn:repl:vs:speedup}. 
\TWOPS{} leads to a similar replication factor as \HDRF{} and \DBH{} on \EN{}. 
However, in contrast to \HDRF{} and \DBH{} which are perfectly balanced, \TWOPS{} leads to a large vertex imbalance resulting in a smaller speedup.

\begin{figure}[t]
\centering
\begin{subfigure}[b]{0.99\linewidth}
\centering
\includegraphics[width=\linewidth]{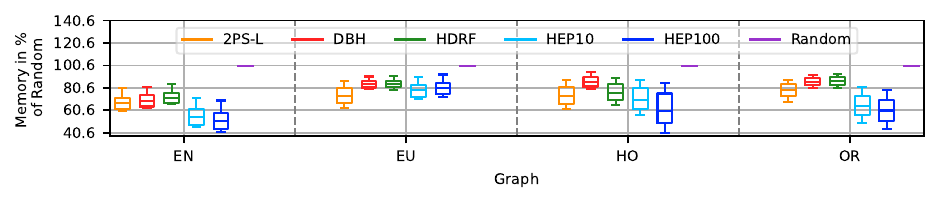}
\caption{4 machines.}
\label{fig:distgnn:memory:4}
\end{subfigure}
\centering
\begin{subfigure}[b]{0.99\linewidth}
\centering
\includegraphics[width=\linewidth]{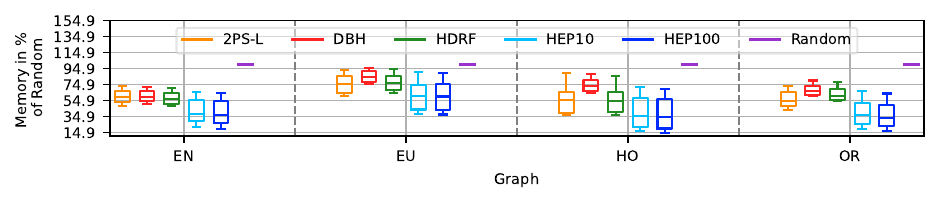}
\caption{32 machines.}
\label{fig:distgnn:memory:32}
\end{subfigure}
\caption{Distribution of memory footprint of graph partitioners in \% of random partitioning on 4 and 32 machines.}
\label{fig:distgnn:memory}
\end{figure}

\blue{Figure \ref{fig:distgnn:memory} gives an overview of how much memory is needed for training with different partitioners and graphs in percent of random partitioning for all GNN parameter combinations \textit{feature size}, \textit{hidden dimension}, and \textit{number of layers}.   
We make two main observations:
(1) The high-quality partitioners (\HEPP{} and \HEPPP{}) are more effective in reducing memory footprint than the other partitioners.
(2) For each combination of graph partitioner and graph in Figure \ref{fig:distgnn:memory}, we observe large deviations, indicating that the partitioners' effectiveness depends on the GNN parameters. 
In the following, we first discuss why the partitioners differ in their effectiveness in terms of memory footprint (observation 1) and then analyze how the partitioners' effectiveness depends on the GNN parameters (observation 2)}.

We observe a strong correlation ($R^2 \geq 0.99$) between the replication factor and memory footprint. Further, we observe that the memory footprint can heavily decrease, e.g., \HEPPP{} reduces the memory for the graphs \EU{}, \OR{}, \HO{}, \EN{}  by 37\% 53\%, 56\% and 60\% on 8 machines, by  44\%, 60\%, 65\% and 63\% on 16 machines and by 40\%, 67\%, 66\% and 63\% on 32 machines, respectively compared to \RANDOM{}. 
There are also cases where random partitioning leads to out-of-memory errors. For example, in many cases, \DI{} can not be processed if random partitioning is applied, but in contrast, the more advanced partitioners enable the processing in such cases. 
In classical distributed graph processing, the replication factor is often minimized with the primary goal of reducing network communication. 
The memory load is less critical, especially if the vertex state is small which is the case for many graph processing algorithms such as Connected Components, PageRank, and K-cores.
In contrast, in GNNs, the vertex state consists of large feature vectors meaning that the vertex state, not the graph structure, dominates the required memory. 
We conclude that \textbf{minimizing the replication factor is crucial for minimizing the memory overhead and can be decisive for \mbox{GNN training.}}

In the following, we analyze how the different GNN parameters influence the effectiveness of the partitioners \textbf{in terms \mbox{of memory}}.  

\begin{figure}[t]
\centering
\begin{subfigure}[b]{0.99\linewidth}
\centering
\includegraphics[width=\linewidth]{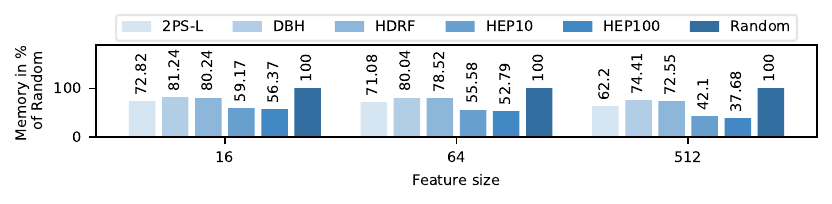}
\caption{Feature size.}
\label{fig:distgnn:memory:feature}
\end{subfigure}
\centering
\begin{subfigure}[b]{0.99\linewidth}
\centering
\includegraphics[width=\linewidth]{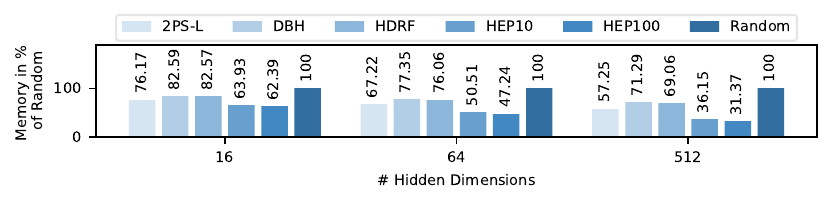}
\caption{Hidden dimensions.}
\label{fig:distgnn:memory:hidden}
\end{subfigure}
\centering
\begin{subfigure}[b]{0.99\linewidth}
\centering
\includegraphics[width=\linewidth]{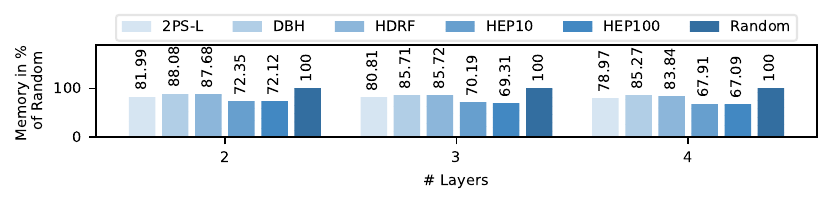}
\caption{Number of layers. Hidden dimension of 16.}
\label{fig:distgnn:memory:layer2}
\end{subfigure}
\centering
\begin{subfigure}[b]{0.99\linewidth}
\centering
\includegraphics[width=\linewidth]{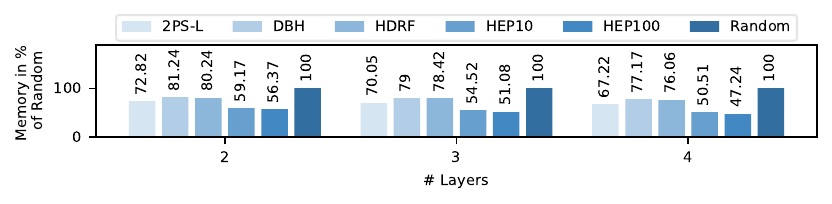}
\caption{Number of layers. Hidden dimension of 64.}
\label{fig:distgnn:memory:layer1}
\end{subfigure}
\caption{Memory footprint in \% of random partitioning dependant on different GNN parameters for \OR{} on 8 machines.}
\label{fig:distgnn:overview-memory}
\end{figure}

\textit{(1) Feature size.} In Figure~\ref{fig:distgnn:memory:feature}, we report the memory footprint for all partitioners in percent of random partitioning dependent on the feature size. 
We observe that if we keep all other parameters constant, \blue{graph partitioning is more effective in the face of larger feature sizes.} 
For example, \HEP10{} has a memory footprint of 59.17\% of \RANDOM{} for a feature size of 16, and a memory footprint of only 42.1\% of \RANDOM{} for a higher feature size of 512. 
This result seems plausible. 
A fixed amount of memory is needed, e.g., for storing the graph structure which is not dependant on the feature size for all partitioners. 
In addition, the features need to be stored, and with an increasing feature size, the memory footprint for storing the features increases. Therefore, the better the partitioning (in terms of less replications of feature vectors), the higher the memory savings. 
\blue{We conclude that \textbf{graph partitioning is more effective in reducing the memory footprint in the face of larger feature sizes.}}

\textit{(2) Hidden dimension.} If the feature size and number of layers are kept constant, there will be some fixed amount of memory for the graph structure, the corresponding features, and the replication of the features. 
We observe that the higher the hidden dimension, the more effective the partitioners become (see Figure~\ref{fig:distgnn:memory:hidden}). 
For example, \TWOPS{} has memory footprint of 76.17\% and 57.25\% of \RANDOM{} for a hidden dimension of 16 and 512, respectively.  
This observation seems plausible.
Larger hidden sizes lead to larger intermediate representations which need to be replicated to other machines as they are needed as the input of the next layer.
\blue{We conclude that \textbf{graph partitioning is more effective in reducing the memory footprint in the face of larger hidden dimensions.}}

\textit{(3) Number of layers.}
Our experiments show that the number of layers can influence the effectiveness of graph partitioning.
The higher the number of layers, the more intermediate representations need to be stored (one per vertex and layer) which are needed in the backward pass. 
The hidden dimension influences the size of the intermediate representations. 
The higher the replication factor, the higher the number of machines to which the intermediate representations are replicated. 
\blue{We observe that graph partitioning becomes more effective as the number of layers increases.}
However, there is also an interesting reinforcing effect between the number of layers and the size of the hidden dimension: With a large hidden dimension, the memory saving of graph partitioning under a growing number of layers becomes even more significant. 
For example, the memory footprint of \HEPPP{} in percent of random partitioning 
is reduced from 72.12\% for 2 layers to 67.09\% for 4 layers if the hidden dimension is 16 (See. Figure~\ref{fig:distgnn:memory:layer2}) and is reduced from 56.37\% for 2 layers to 47.24\% for 4 layers if the hidden dimension is 64 (See. Figure~\ref{fig:distgnn:memory:layer1}). 
We conclude, \textbf{the more layers, the more important \mbox{the partitioning.}}

\begin{figure}[t]
\centering
\begin{subfigure}[b]{0.99\linewidth}
\centering
\includegraphics[width=\linewidth]{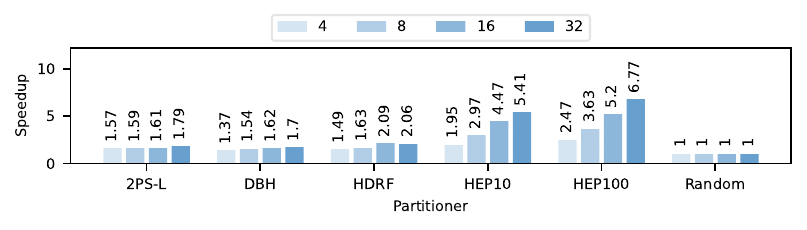}
\caption{Speedup (larger is better).}
\label{fig:distgnn:scaleout:speedup}
\end{subfigure}
\centering
\begin{subfigure}[b]{0.99\linewidth}
\centering
\includegraphics[width=\linewidth]{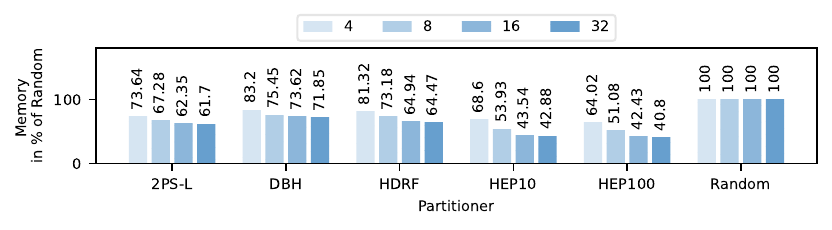}
\caption{Memory overhead (lower is better).}
\label{fig:distgnn:scaleout:memory}
\end{subfigure}
\begin{subfigure}[b]{0.99\linewidth}
\centering
\includegraphics[width=\linewidth]{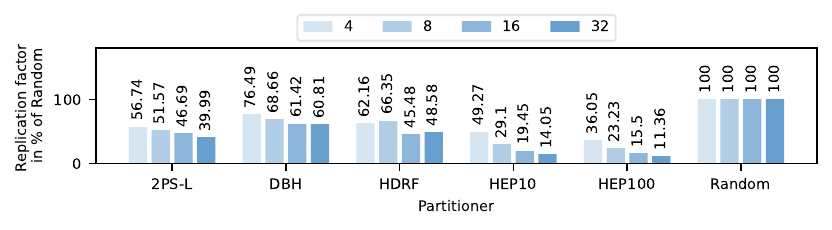}
\caption{Replication factor (lower is better).}
\label{fig:distgnn:scaleout:replicationfactor}
\end{subfigure}
\caption{The effectiveness of graph partitioning increases in terms of speedup and memory overhead if the scale-out factor is increased from 4 to 8, 16 and 32 machines.}
\label{fig:distgnn:scaleout}
\end{figure}

\textit{(4) Scale-out factor.}
\blue{In the following, we analyze the effectiveness of graph partitioning in terms of speedup and memory footprint across various scale-out factors.}
Figure \ref{fig:distgnn:scaleout:speedup} shows the average speedup for all graph partitioners at different scale-out factors compared to random partitioning.  
We observe that the effectiveness of all graph partitioners increases if more machines are used for training. 
However, there are differences in how sharply the effectiveness increases. 
For the more light-weight partitioners \TWOPS{}, \DBH{} and \HDRF{}, the speedups increase moderately from 1.57x, 1.37x, and 1.49x on 4 machines to 1.79x, 1.7x, and 2.06x on 32 machines, respectively. The partitioners \HEPP{} and \HEPPP{} lead to a speedup of 1.95x and 2.47x on 4 machines which increase sharply to 5.41x and 6.77x on 32 machines, respectively. 
We make similar observations for the memory overhead (see Figure~\ref{fig:distgnn:scaleout:memory}). 
All partitioners lead to substantial savings which increase if the number of machines increases. 
Both observations are plausible:
In Figure \ref{fig:distgnn:scaleout:replicationfactor}, we report the achieved replication factors for all partitioners for all scale-out factors in percentages of \RANDOM{}, meaning lower numbers are better. We find that the replication factor of all partitioners increases less sharply than \RANDOM{} if the scale-out factor increases. 
We observe that the replication factors of \TWOPS{}, \DBH{} and \HDRF{} are 56.74\%, 76.49\% and 62.16\% of \RANDOM{} on 4 machines and 39.99\%, 60.81\% and 48.58\% of \RANDOM{} on 32 machines. \HEPP{} and \HEPPP{} achieve a replication factor of 49.27\% and 36.05\% of \RANDOM{} on 4 machines and significantly lower replication factors of 14.05\% and 11.37\% of \RANDOM{} on 32 machines, respectively.  
\blue{We conclude that \textbf{graph partitioning is more effective in reducing training time and memory footprint in the face of larger scale-out factors.}}

\textit{(5) Partitioning time amortization.}
In Table~\ref{tab:amortization}, we report the average number of epochs until the partitioning time is amortized by faster training time for each combination of graph and partitioner. 
We assume that random partitioning does not take any time.
\DBH{} is the partitioner that amortizes the fastest: on average, it takes 1.39, 3.79, 3.05, and 3.83 epochs on the graphs \EN{}, \EU{}, \HO{} and \OR{} to amortize the partitioning time. \HEPPP{} which leads to the largest speedups amortizes after 4.29, 12.0, 4.7, and 7.03 epochs on the graphs \EN{}, \EU{}, \HO{} and \OR{}. Training is often performed for hundreds of epochs \cite{DistGNN}, \textbf{therefore, the partitioning time can be amortized}. In addition, a hyper-parameter search is often performed which requires even more training epochs. Therefore, it is even more beneficial to invest time into partitioning.

\begin{table}[t]
\caption{Number of epochs until amortization.}
\label{tab:amortization}
\begin{center}
\begin{tabular}{|l||l|l|l|l|l|}
\hline
Graph & DBH & 2PS-L & HDRF & HEP10 & HEP100\\ 
\hline \hline
EN & 1.39 & 4.57 & 4.64 & 3.35 & 4.29\\ 
 \hline
EU & 3.79 & never  & 8.8 & 10.15 & 12.0\\ 
 \hline
HO & 3.05 & 4.22 & 7.26 & 4.48 & 4.7\\ 
 \hline
OR & 3.83 & 7.39 & 11.69 & 6.64 & 7.03\\ 
 \hline
\end{tabular}
\end{center}
\end{table}

%% file: sections/distdgl_results.tex
\section{DistDGL - Vertex Partitioning }
\label{sec:results:distdgl}

\subsection{Experiments}
\label{sec:distdgl:experiments}
\textbf{Graph partitioning algorithms.} 
 We use six state-of-the-art vertex partitioners from different categories: 
(1)~stateless streaming partitioning with random partitioning, 
(2)~stateful streaming with LDG~\cite{ldg}, and 
(3)~in-memory partitioning with ByteGNN~\cite{bytegnn}, Spinner~\cite{spinner}, Metis~\cite{metis}, and KaHIP~\cite{sandersschulz2013}.

\noindent\textbf{Workloads.} We selected a representative set of graph neural network architectures commonly used in distributed GNN training, namely, GAT, GraphSage, and GCN. 
We use the same hyper\-pa\-ra\-me\-ters as for \textit{DistGNN} (see Table~\ref{tab:hyper-parameters}). 
\DISTDGL{} uses mini-batch training with neighborhood sampling. 
If not mentioned otherwise, we perform neighborhood sampling for all GNN models with the following configuration.
Let $S_i$ be the number of neighbors to sample for layer~$i$. 
For two layer GNNs, we use $l_1=25$ and $l_2=20$, for three layer GNNs $l_1=15$, $l_2=10$ and $l_3=5$ and for four layer GNNs, $l_1=10$, $l_2=10$, $l_3=5$ and $l_4=5$. 
We use a global batch size $GBS$ of 1024 if not stated otherwise. 
Therefore, each worker $w_i \in W$ trains with $\frac{GBS}{|W|}$ samples. 

\noindent\textbf{Partitioning metrics.}
We compare the partitioners with the commonly used partitioning quality metrics \textit{edge-cut} and \textit{vertex balance} introduced in Section~\ref{sec:background:partitioning}. 
In addition, we measure the training vertex balance. 
Further, we measure metrics based on the sampled mini-batches: the number of edges of the computation graphs, the number of local input vertices, and the number of vertices that need to be fetched via the network. 

\noindent\textbf{Training \& Partitioning Time.}
We measure the epoch and step time and all phases (mini-batch sampling, feature loading, forward pass, backward pass, and model update) for each step.
In addition, we measure the partitioning time. 

\subsection{Partitioning Performance}

\label{distdgl:part-metrics}
\begin{figure}[t]
\centering
\begin{subfigure}[b]{0.99\linewidth}
\centering
\includegraphics[width=\linewidth]{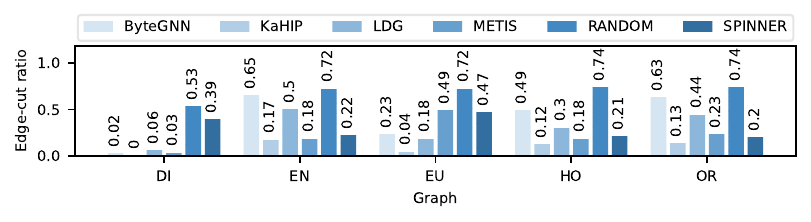}
\caption{4 partitions.}
\label{fig:edge-cut:4}
\end{subfigure}
\begin{subfigure}[b]{0.99\linewidth}
\centering
\includegraphics[width=\linewidth]{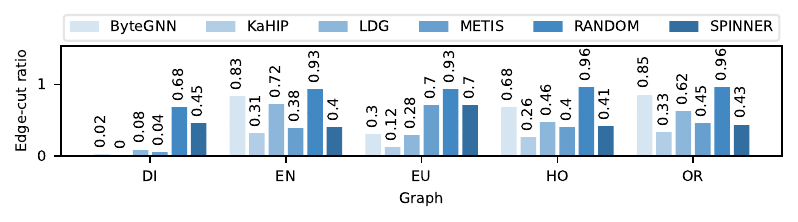}
\caption{32 partitions.}
\label{fig:edge-cut:32}
\end{subfigure} 
\caption{Edge-cut ratio for all combinations of graph, number of partitions and graph partitioner.}
\label{fig:edge-cut}
\end{figure} 
 
In the following, we compare the graph partitioners regarding communication costs and computational balance in \mbox{GNN training.}

\textit{(1) Communication Costs.}
In Figure~\ref{fig:edge-cut}, we report the edge-cut ratio obtained for each combination of graph, graph partitioning algorithm, and number of partitions. 
In most cases, \KAHIP{} achieves the lowest edge-cut, and random partitioning leads to the largest edge-cut. 
We observe significant differences between the partitioning algorithms in terms of edge-cut, e.g., \KAHIP{} achieves an edge-cut ratio smaller than 0.001 and 0.12 on the graph \DI{} and \EU{} for 32 partitions, respectively, which is much lower (better) compared to random partitioning which leads to edge-cuts of 0.68 and 0.93 for the same graphs. 
We also observed for all partitioning algorithms that a higher number of partitions leads to a larger edge-cut. 

In the following, we investigate the influence of the edge-cut on network communication. 
There are cases where a lower edge-cut results in less network communication. 
However, there are also cases where even if the edge-cut of different partitioners is similar, the network communication can differ a lot. 
For example, we observed that Spinner has an edge-cut lower than Metis on \OR{}, but the network communication is higher.  
This observation can be explained as follows. 
There can be edges that are more frequently involved in the sampling process. 
If these edges are cut, they lead to more network traffic than edges hardly visited in the sampling process. 
To ensure that the observed anomaly is related to graph partitioning, we measure for each mini-batch the number of vertices needed for processing the mini-batch that are not local to the respective worker. 
We define these vertices as \textit{remote vertices}.
We observe a strong correlation between the number of remote vertices and network traffic. 
We conclude that edge-cut is not always a perfect predictor for network traffic and that there are cases where a lower edge-cut still leads to more remote vertices and higher network traffic. 

\begin{figure}[t]
\centering
\begin{subfigure}[b]{0.99\linewidth}
\centering
\includegraphics[width=\linewidth]{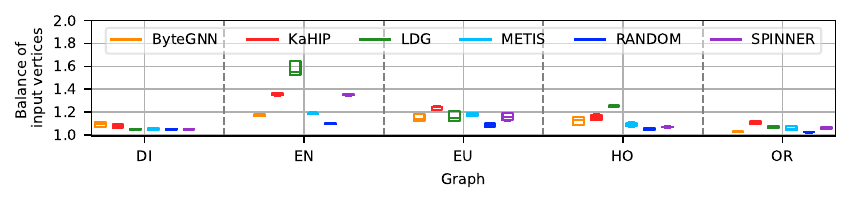}
\caption{Input vertex balance.}
\label{fig:balancing:minibatch:8}
\end{subfigure} 
\begin{subfigure}[b]{0.99\linewidth}
\centering
\includegraphics[width=\linewidth]{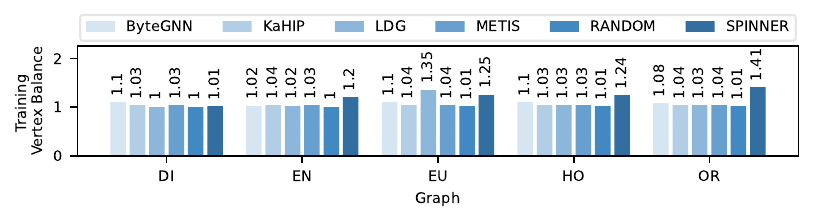}
\caption{Training vertex balance.}
\label{fig:balancing:training_vertices:8}
\end{subfigure}
\caption{Different balancing metrics for 8 partitions.}
\label{fig:dgl:balancing:partitioning}
\end{figure}

\textit{(2) Computation Balance.}
For efficient distributed graph processing, it is crucial that the computation is balanced among machines which is commonly associated with the vertex balance.
However, unlike distributed graph processing algorithms such as PageRank, the computational load of mini-batch GNN training depends on the size of the sampled mini-batches. 
Each worker samples a mini-batch based on the k-hop neighborhood of the training vertices. 
To ensure load balance, it is essential that the computation graphs of mini-batches are of similar size. 
We define the number of vertices that are needed to compute a mini-batch as the \textit{input vertices} and \textit{input vertex balance} per step as the number of input vertices of the largest mini-batch divided by the average number of input vertices per mini-batch in the respective step.
In Figure~\ref{fig:balancing:minibatch:8}, we report the imbalance of the mini-batches in terms of input vertices.
We observe a large imbalance, which is increasing as the number of partitions is increased.
This imbalance occurs, although the number of training vertices is balanced (see. Figure~\ref{fig:balancing:training_vertices:8}).
This observation indicates that tailoring the partitioning to the sampling and vice versa could be an interesting research direction to ensure computational balancing. 

\begin{figure}[t]
\centering
\begin{subfigure}[b]{0.99\linewidth}
\centering
\includegraphics[width=\linewidth]{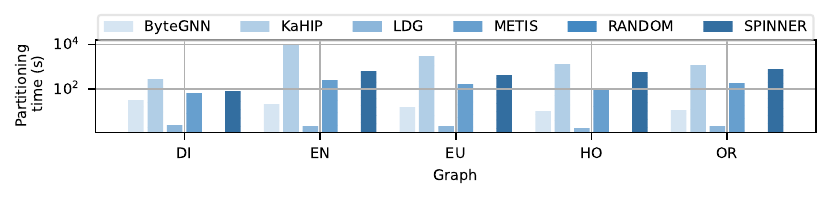}
\caption{4 partitions.}
\label{fig:edge-cut:4}
\end{subfigure}
\hfill
\begin{subfigure}[b]{0.99\linewidth}
\centering
\includegraphics[width=\linewidth]{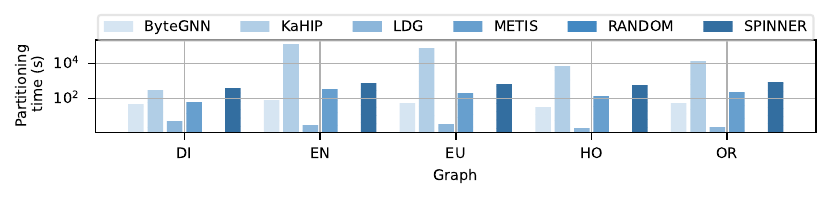}
\caption{32 partitions.}
\label{fig:dist:dgl:partitioning-time:32}
\end{subfigure} 
\caption{Partitioning time on a logarithmic scale.}
\label{fig:dist:dgl:partitioning-time}
\end{figure}

\textit{(3) Partitioning Time.}
In Figure~\ref{fig:dist:dgl:partitioning-time}, we report the partitioning time for all graphs and partitioning algorithms for 4 and 32 partitions. We observe that the best (in terms of lowest edge-cut) performing partitioner \KAHIP{}, leads to the highest \mbox{partitioning time.}

\subsection{GNN Training Performance}
\label{sec:distdgl:influence-gnn-parameters}
\begin{figure}[t]
\centering
\begin{subfigure}[b]{0.99\linewidth}
\centering
\includegraphics[width=\linewidth]{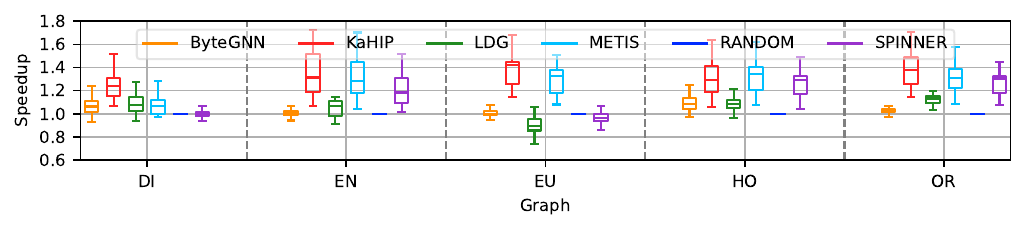}
\caption{4 machines.}
\label{ig:overview-epoch-times:4}
\end{subfigure}
\hfill
\begin{subfigure}[b]{0.99\linewidth}
\centering
\includegraphics[width=\linewidth]{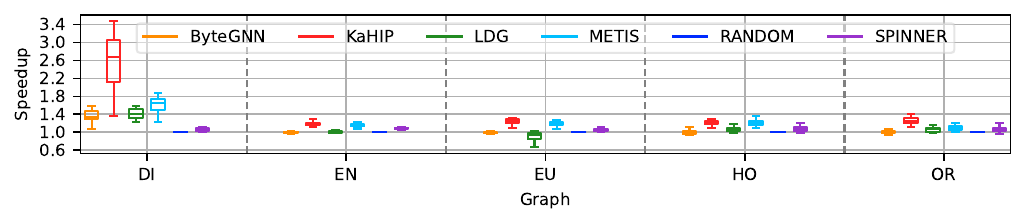}
\caption{32 machines.}
\label{fig:overview-epoch-times:32}
\end{subfigure}
\hfill
\caption{Speedup distribution of graph partitioners on 4 and 32 machines for all GraphSage experiments.}
\label{fig:overview-epoch-times}
\end{figure}

\begin{figure}[t]
\centering
\begin{subfigure}[b]{0.99\linewidth}
\centering
\includegraphics[width=\linewidth]{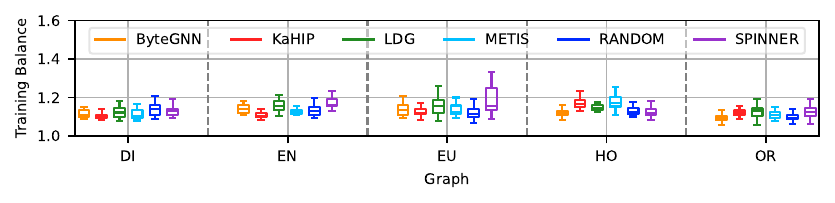}
\caption{4 machines.}
\label{fig:balancing:sum_phases:4}
\end{subfigure}
\begin{subfigure}[b]{0.99\linewidth}
\centering
\includegraphics[width=\linewidth]{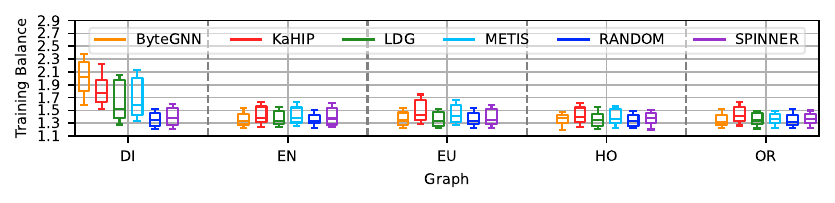}
\caption{32 machines.}
\label{fig:balancing:sum_phases:32}
\end{subfigure} 
\caption{Balance in terms of training time.}
\label{fig:balance:phase-times}
\end{figure}

In Figure~\ref{fig:overview-epoch-times}, we report the average speedups for all combinations of \textit{feature size}, \textit{hidden dimension} and \textit{number of layers} (see Table~\ref{tab:hyper-parameters}) for all graph partitioners with random partitioning as a baseline on 4 and 32 machines for the GraphSage \cite{hamilton2017inductive} architecture. 
In our experiments, \KAHIP{} and \METIS{} lead to the largest speedups of up to 1.84x, 1.84x, 3.09x and 3.47x on a cluster with 4, 8, 16 and 32 machines, respectively. 
We see significant variances in terms of speed up that indicate that the effectiveness of the partitioning algorithms depends on the GNN parameters. Therefore, we conduct a detailed analysis of how different GNN parameters influence the different training phases and how the partitioners differ from each other. 
On each worker in each step, we measure the phase times (1) mini-batch sampling, (2) feature loading, (3) forward pass, (4) backward pass, and (5) model update. 
In each step, we identify the worker that leads to the longest mini-batch sampling, feature loading and forward pass time as the straggler. 
We exclude the time for the backward pass because it also contains the time for the all-reduce operation in which the gradients are synchronized between the workers.
The model update time is negligible. 
Then, we get the phase times of the slowest worker per step and sum up the phase times of all steps of the respective slowest worker. 
In other words, we are interested in how much time the straggler spends on average in each phase.
\blue{In Figure~\ref{fig:balance:phase-times}, we report the imbalances in terms of training time for all combinations of GNN parameters per graph and graph partitioner. 
We observe imbalances (values greater~1) for all partitioners and graphs, showing that even if the number of training vertices is balanced, training time can be imbalanced.} 
As discussed in Section~\ref{distdgl:part-metrics}, this can be attributed to an imbalance in the computation graphs which is caused by the sampling step.
In the following, we investigate how the different GNN model parameters influence the effectiveness of graph partitioning in terms of speedup of the distributed training compared to \mbox{random partitioning.}

\begin{figure}[t]
\centering
\begin{subfigure}[b]{0.97\linewidth}
\centering
\includegraphics[width=\linewidth]{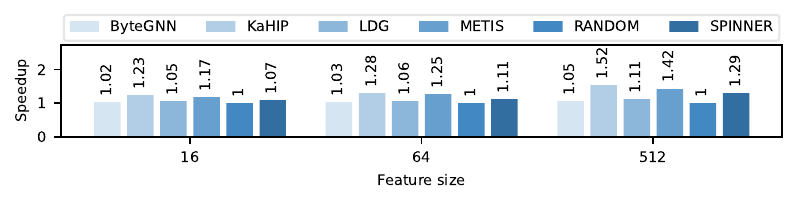}
\caption{Feature size.}
\label{fig:overview-epoch-times:feature-size:4}
\end{subfigure}
\begin{subfigure}[b]{0.97\linewidth}
\centering
\includegraphics[width=\linewidth]{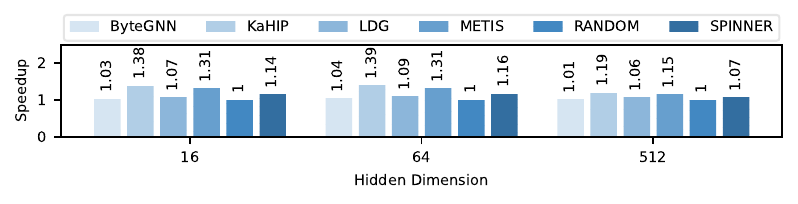}
\caption{Hidden dimension.}
\label{fig:overview-epoch-times:hidden-dim:4}
\end{subfigure}
\begin{subfigure}[b]{0.97\linewidth}
\centering
\includegraphics[width=\linewidth]{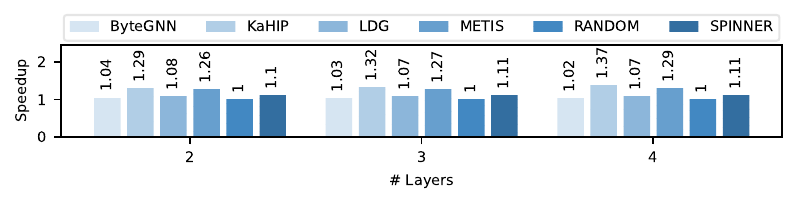}
\caption{Number of Layers.}
\label{fig:overview-epoch-times:num-layers:4}
\end{subfigure}
\caption{Speedup of graph partitioners for the GraphSage model on 4 machines for different GNN parameters.}
\label{fig:dgl:speedups:params}
\end{figure}

\blue{\textit{(1) Feature size.} We observe that \textbf{graph partitioning is more effective in the face of larger feature sizes.}}
Figure~\ref{fig:overview-epoch-times:feature-size:4} shows the speedups for the partitioners compared to random partitioning dependent on the feature size.
For example, the training for GraphSage with \KAHIP{} leads to a speedup of 1.23x and 1.52x for a feature size of 16 and 512, respectively.  
This observation is plausible. 
As feature sizes increase, network communication increases because larger feature vectors are sent over the network, making graph partitioning even more valuable in reducing communication costs.

\begin{figure}[t]
\centering
\includegraphics[width=0.95\columnwidth]{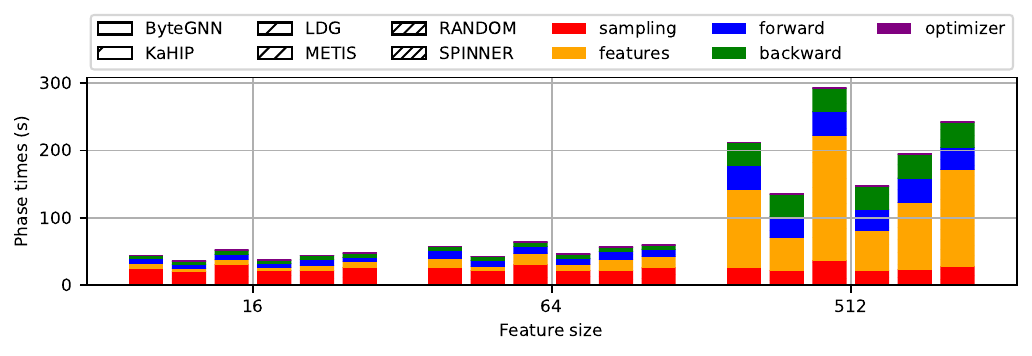}
\caption{Phase times for a 3 layer GraphSAGE with a hidden dimension of 64 on 4 machines \mbox{on \EU{} graph.}}   
\label{fig:overview:phases:eu-2015-tpd}
\end{figure}

For each combination of number of layers and hidden dimension, we vary the feature size. 
We make the following key observations: (1) The larger the feature size, the longer the feature fetching phase (see Figure~\ref{fig:overview:phases:eu-2015-tpd}) while the sampling time stays constant. We also observe that for small feature sizes (up to 64), the sampling takes more time than fetching features, but for large feature sizes of 512, the time for feature fetching dominates the sampling time by a lot (see Figure \ref{fig:overview:phases:eu-2015-tpd}). 
Only, for the road network \DI{}, we observe that sampling always takes more time than fetching features which seems plausible because the mean degree in the road network is small and the skew of the degree distribution is low. Therefore, the sampled mini-batches are small, and only a few input vertices must be fetched. We also observe that the edge-cut for the road network is much lower than for the remaining graphs (see Figure~\ref{fig:edge-cut}). (2) The forward and backward pass time increases with larger feature sizes, which is plausible because more computations will be performed in the first layer.

We observe that the partitioners differ a lot from each other when varying the feature size and that the feature size influences the different phases. 
The feature fetching phase is influenced most, which can for example be seen in Figure~\ref{fig:overview:phases:eu-2015-tpd} for training a three layer GraphSage with a hidden dimension of 64 on the graph \EU{}. In most cases, the better the partitioner in terms of edge-cut, the lower the communication costs, which can speed up both the mini-batch sampling and the feature fetching phase.

\textit{(2) Number of hidden dimensions.}
\label{sec:num-hidden}
We found that \textbf{partitioning becomes less crucial as the hidden dimension increases}.
In Figure~\ref{fig:overview-epoch-times:hidden-dim:4} we report the speedups for the partitioners compared to random partitioning dependent on the number of hidden dimensions.
For example, compared with random partitioning, \KAHIP{} leads to a speedup of 1.38x and 1.19x, and \METIS{} to 1.31x and 1.15x  for a hidden dimension of 16 and 512, respectively. 
This result is reasonable since an increased hidden dimension leads to greater computational costs, potentially dominating the communication costs. 
We vary the hidden dimension for each combination of feature size and number of layers. 
Our main observations are: 
(1)~sampling and feature loading time stay constant, which is expected as only the neural network operations are influenced by the hidden dimension. 
The larger the hidden dimension, the more time is used for computation.
(2) We also observe that the effectiveness of partitioners decreases for the larger hidden dimension because most of the differences are in feature loading and sampling. However, if the hidden dimension increases, the computation takes a larger share of the overall training time. 
Therefore, the difference between the partitioners is lower.

\begin{figure}[t]
\centering
\includegraphics[width=0.95\columnwidth]{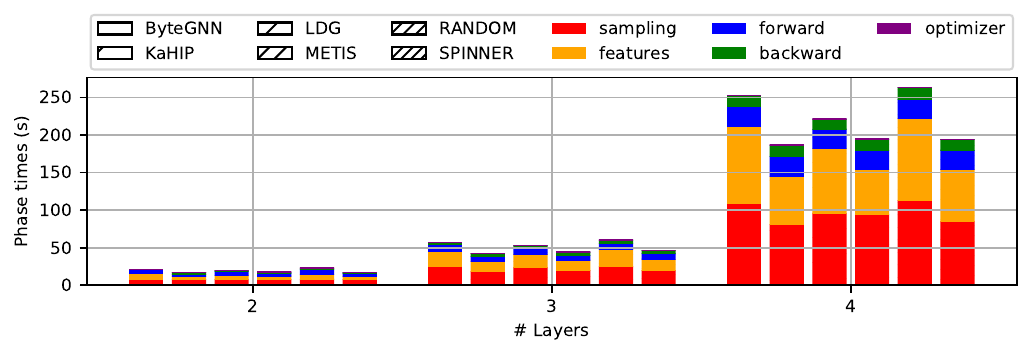}
\caption{Phase times for a GraphSAGE with a hidden dimension and feature size of 64 on 4 machines on \OR{} graph.} 
\label{fig:overview:phases:layer}
\end{figure}

\begin{figure}[t]
\centering
\includegraphics[width=0.95\columnwidth]{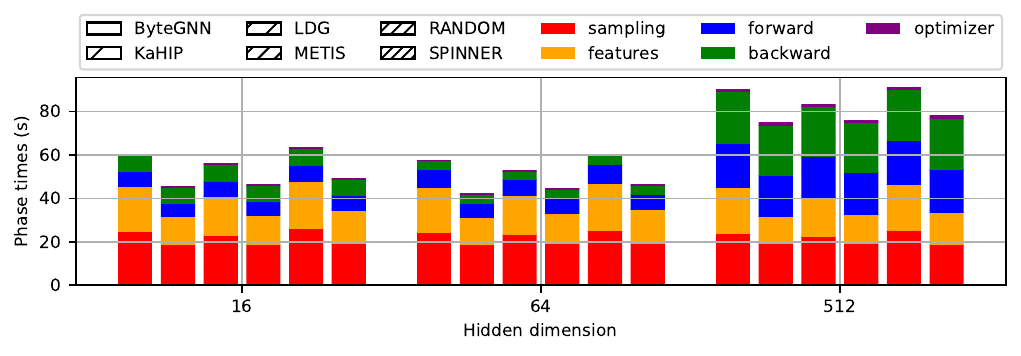}
\caption{Phase times for a 3 layer GraphSAGE with a feature size of 64 on 4 machines on \OR{} graph.}    
\label{fig:overview:phases:hidden}
\end{figure}

\textit{(3) Number of Layers.}
We observe that \textbf{the effectiveness of graph partitioning remains relatively unaffected by an increasing number of layers}. 
Figure~\ref{fig:overview-epoch-times:num-layers:4} plots the speedups for the partitioners compared to random partitioning dependent on the number of layers.
In some cases, the effectiveness slightly increases or decreases, but the influence is much smaller than the influence of the feature size and hidden dimension, and there is also no clear trend.
This is an unexpected observation. 
One could think that the effectiveness of the partitioning algorithms would heavily decrease if the number of layers increases because large parts of the graph will be contained in the mini-batches, but still, the partitioning algorithms lead to different training times, and many partitioners outperform random partitioning.
When varying the number of layers for each combination of feature size and hidden dimension, we make the following key observations: (1) All phases increase in run-time if the number of layers increases. This is expected because an increase in the number of layers leads to larger computation graphs within the mini-batches, which increase the communication costs (more remote accesses in the sampling phase and more remote vertices to fetch via the network) and the computation costs (more neural network operations). (2) We observe, especially for 3 and 4 layer GraphSage, that most of the speedup gained by different partitioning algorithms comes from faster sampling and feature fetching (see Figure \ref{fig:overview:phases:layer}). 

\begin{figure}[t]
\centering
\begin{subfigure}[b]{0.95\linewidth}
\centering
\includegraphics[width=\linewidth]{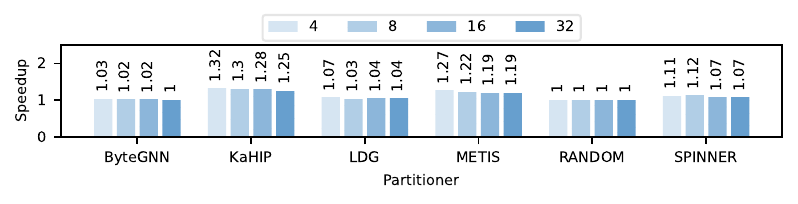}
\caption{Speedup.}
\label{fig:scaling:speedup}
\end{subfigure}
\hfill
\begin{subfigure}[b]{0.95\linewidth}
\centering
\includegraphics[width=\linewidth]{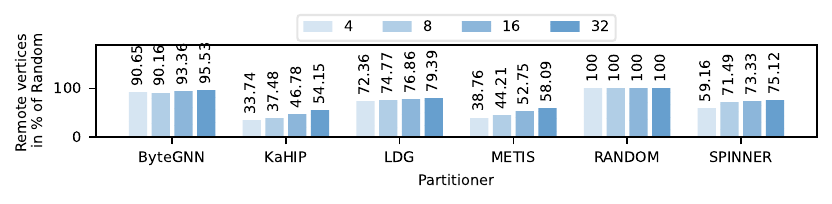}
\caption{Remote vertices.}
\label{fig:scaling:remotevertices}
\end{subfigure}
\hfill
\begin{subfigure}[b]{0.95\linewidth}
\centering
\includegraphics[width=\linewidth]{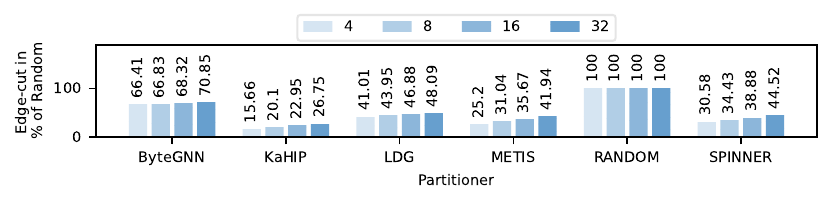}
\caption{Edge-cut.}
\label{fig:scaling:edgecut}
\end{subfigure}
\hfill
\begin{subfigure}[b]{0.95\linewidth}
\centering
\includegraphics[width=\linewidth]{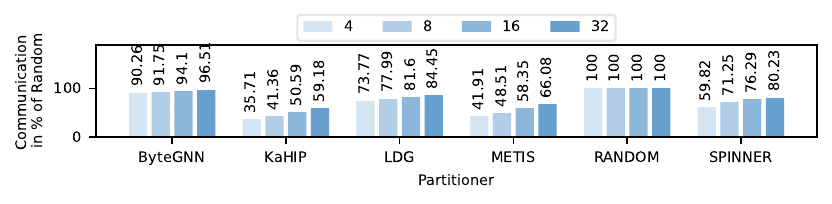}
\caption{Communication.}
\label{fig:scaling:network}
\end{subfigure}
\hfill
\begin{subfigure}[b]{0.95\linewidth}
\centering
\includegraphics[width=\linewidth]{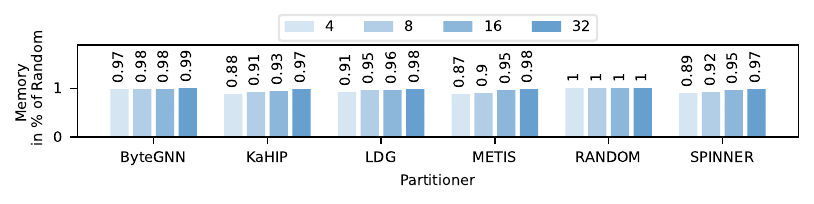}
\caption{Memory.}
\label{fig:scaling:memory}
\end{subfigure}
\caption{The effectiveness of partitioning decreases when scaling GraphSage from 4 to 32 machines.}
\label{fig:scaling:overview}
\end{figure}

\begin{figure}[t]
\centering
\includegraphics[width=0.95\columnwidth]{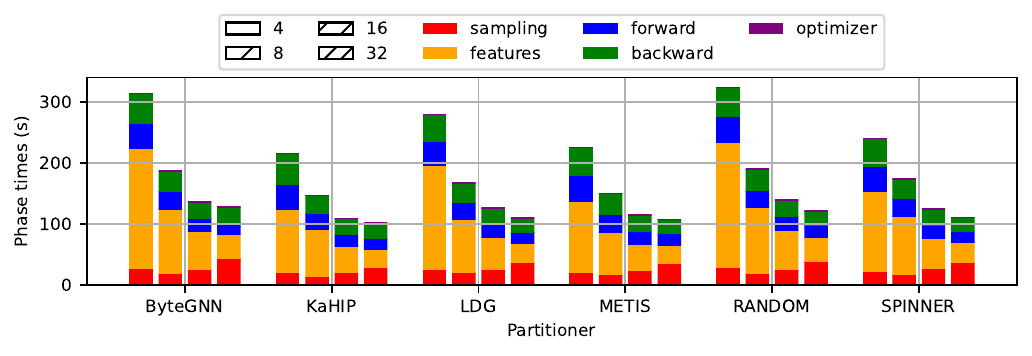}
\caption{Phase times for a 3 layer GraphSage with a feature size of 512 and a hidden dimension of 64 on \OR{} graph trained with 4, 8, 16 and 32 machines.}    
\label{fig:scaling:phases}
\end{figure}

\textit{(4) Scale-out factor.}
\label{sec:distdgl:scale-out}
\blue{In the following, we investigate the effectiveness of graph partitioning in the face of larger scale-out factors.}
We scale out from 4 to 8, 16 and 32 machines. 
We make the following observations: 
(1) For \DI{}, scaling out increases the effectiveness of the partitioners. 
Especially for the partitioners \KAHIP{}, \METIS{}, \LDG{} and \BYTEGNN{}, the effectiveness increases a lot.
However, for \SPINNER{}, the effectiveness stays relatively constant.
This seems plausible as the edge-cut for \RANDOM{} and \SPINNER{} is far higher on \DI{} than for the remaining partitioners (see Figure~\ref{fig:edge-cut}).
(2) For the remaining graphs, we observe that the effectiveness on GraphSage decreases on average (see Figure~\ref{fig:scaling:speedup}). 
For example, \KAHIP{} and \METIS{} lead to a speedup of 1.32x and 1.27x on 4 machines and to a smaller speedup of 1.25x and 1.19x on 32 machines, respectively. 
We found that the number of remote vertices (see Figure~\ref{fig:scaling:remotevertices}) and the edge-cut (see Figure~\ref{fig:scaling:edgecut}) of the partitioners in percentages of \RANDOM{} is increasing when scaling out to more machines. 
We also observe that the network communication of the partitioners in percentages of \RANDOM{} is also increasing. In other words, the effectiveness of the partitioners is also decreasing in terms of partitioning metrics and network communication compared to \RANDOM{}, when the number of machines increases. 
It is worth to note that the feature loading phase scales well. 
\blue{We found that graph partitioning is more effective in the face of large feature vectors.}
For large feature vectors and few machines, the feature fetching phase can take a large share of the training time and also leads to large differences between the partitioners (see Figure~\ref{fig:scaling:phases}). 
The feature fetching phase can decrease sharply when scaling out to more machines. Therefore, the difference between the partitioners is decreasing, resulting in lower effectiveness. We make a similar observation with the sampling phase. 
Further, we observe that the effectiveness in terms of memory is decreasing (see Figure~\ref{fig:scaling:memory}) with more machines. This is plausible. For each mini-batch, feature vectors of remote vertices need to be fetched and stored. Therefore, each remote vertex leads to some additional memory. As mentioned above, the number of remote vertices in percent of random partitioning increases with more machines (see Figure~\ref{fig:scaling:remotevertices}) meaning less memory can be saved. 
\blue{We conclude that \textbf{in most cases graph partitioning is slightly less effective in the face of larger scale-out factors.}}

\textit{(5) Partitioning time amortization.}
In Table~\ref{tab:distdgl:amortization}, we report the average number of epochs until the partitioning time is amortized by faster training time for each combination of graph and partitioner. 
We observe that \textbf{the partitioning time can be amortized by faster GNN training}. 
However, \KAHIP, the partitioner which leads to the largest speedups only amortizes for \DI{}, but barely for the remaining graphs. 
\METIS{}, which also leads to significant speedups, does amortize for all graphs within less than 20 epochs.   

\begin{table}[t]
\caption{Number of epochs until amortization.}
\label{tab:distdgl:amortization}
\begin{center}
\begin{tabular}{|l||l|l|l|l|l|}
\hline

Graph & ByteGNN & KaHIP & LDG & SPINNER & METIS\\ 
\hline \hline
DI & 0.93 & 2.61 & 0.1 & 14.37 & 1.13\\ 
 \hline
EN & 2.16 & 2501.93 & 0.39 & 54.07 & 16.79\\ 
 \hline
EU & never & 1197.25 & never & 53.8 & 8.14\\ 
 \hline
HO & 0.68 & 347.51 & 0.47 & 77.78 & 10.7\\ 
 \hline
OR & 3.14 & 223.19 & 0.27 & 70.19 & 14.59\\ 
 \hline
\end{tabular}
\end{center}
\end{table}

\subsection{Influence of mini-batch size}
\label{sec:distdglmini-batch-effectiveness}

\begin{figure}[t]
\begin{subfigure}[b]{0.90\linewidth}
\centering
\includegraphics[width=\linewidth]{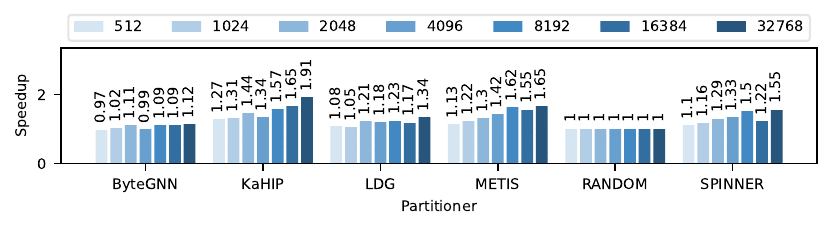}
\caption{Speedup compared to \RANDOM{}.}
\label{fig:vary-mini-batch-size:speedup}
\end{subfigure} 
\hfill
\begin{subfigure}[b]{0.95\linewidth}
\centering
\includegraphics[width=\linewidth]{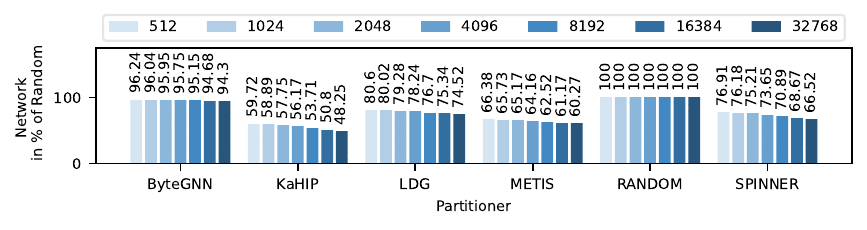}
\caption{Network communication in \% of \RANDOM{}.}
\label{fig:vary-mini-batch-size:network}
\end{subfigure} 
\hfill
\begin{subfigure}[b]{0.95\linewidth}
\centering
\includegraphics[width=\linewidth]{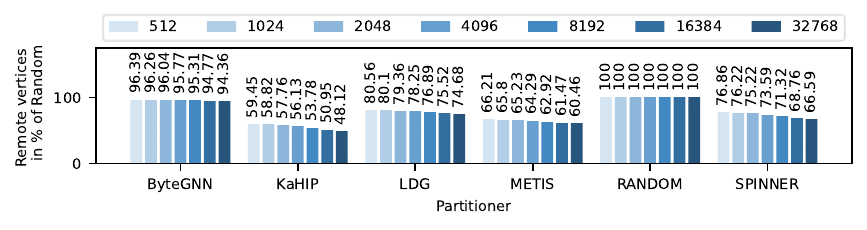}
\caption{Number of remote vertices in \% of \RANDOM{}.}
\label{fig:vary-mini-batch-size:remote-vertices}
\end{subfigure}
\caption{Different metrics for a 3 layer GraphSage with a hidden dimension of 64 and feature size of 512 on the \OR{} graph for varying batch sizes between 512 and 32768.}
\label{fig:vary-mini-batch-size}
\end{figure}

The following experiments aim to investigate the influence of the \textit{mini-batch size} on the effectiveness of partitioning. 
In other words, we want to evaluate if the partitioning is more crucial (in terms of reduced training time) if the mini-batch size increases.
 
We fix the number of workers to 16 and set the mini-batch size to 512, 1024, 2048, 4096, 8192, 16384, and 32768 for a three layer GAT and a three layer GraphSage. 
For both GNN architectures, we use two configurations: 
(1)  hidden dimension and feature size of 64 (low communication) and (2) hidden dimension of 64 and feature size of 512 (high communication).

We observe for all partitioners that the network traffic decreases compared to \RANDOM{} when the batch size increases (see Figure~\ref{fig:vary-mini-batch-size:network}). 
For example, \KAHIP{} and \SPINNER{} lead to a network communication of 66\% and 77\% of \RANDOM{} with a batch size of 512 and of 48\% and 67\% with a batch size of 32768, respectively. 
We observe a similar trend for the number of remote vertices (see Figure~\ref{fig:vary-mini-batch-size:remote-vertices}).
This seems reasonable.
Many vertices can end up in \textit{different} mini-batches. 
However, if the mini-batch increases, the overlap in the larger mini-batches increases, leading to fewer remote vertices.

The effectiveness of the partitioners can decrease or increase with larger batch sizes if the feature size is low (64): 
However, there is no clear trend. 
In contrast, if the feature size is high (512), in most cases, the effectiveness of the partitioners increases. 
For example, Figure~\ref{fig:vary-mini-batch-size:speedup} shows that training with \KAHIP{} and \METIS{} leads to a speedup of 1.27x and 1.13x for a small batch size of 512 and to a larger speed up of 1.91x and 1.65x if the batch size is set to 32768.  
\blue{We conclude that \textbf{graph partitioning is more effective in the face of larger batch and feature sizes.}}

\subsection{\blue{Distributed GPU Training}}
\blue{We conducted distributed training on 8 NVIDIA Jetson AGX Orin GPUs, each equipped with 64~GB GPU memory, resulting in a total of 512~GB of GPU memory. 
We trained the GraphSage model with \DISTDGL{} on all graphs of Table~\ref{tab:datasets}. 
In addition, we integrated the \textit{ogbn-papers100M (PA)} citation graph from the Open Graph Benchmark~(OGB)~\cite{hu2020open} which has 111 million vertices and 1.6 billion edges. 
All partitioners were applied to all graphs, except for \KAHIP{} which could not partition the \PA{} graph\footnote{\blue{\KAHIP{} returns "The graph is too large."}}.}
\blue{The GNN parameters are set to medium values: the \textit{hidden dimension} is set to 64, the \textit{number of layers} to 3, and the \textit{batch size} to 1024 per GPU.
We use two different configurations for the feature size representing low and high communication, respectively.
Low communication: The \textit{feature size} is set to 64 for all graphs.  
High communication: The \textit{feature size} is set to 512 for all graphs except \PA{} which runs out of GPU memory in the training process for a feature size of 512. Therefore, the \textit{feature size} was set to 256 instead of 512 for \PA{}.}
\begin{figure}[t]
\centering
\begin{subfigure}[b]{0.95\linewidth}
\centering
\includegraphics[width=\linewidth]{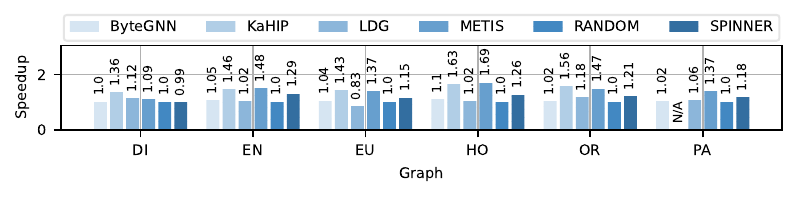}
\caption{\blue{Low communication.}}
\label{fig:distdgl-gpu:low}
\end{subfigure}
\centering
\begin{subfigure}[b]{0.95\linewidth}
\centering
\includegraphics[width=\linewidth]{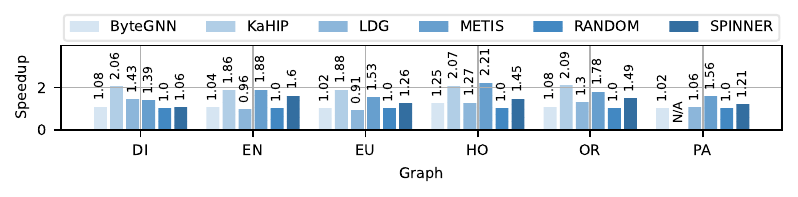}
\caption{\blue{High communication.}}
\label{fig:distdgl-gpu:high}
\end{subfigure}
\caption{\blue{Speedups for distributed GPU training.}}
\label{fig:distdgl-gpu:speedup}
\end{figure}

\blue{Figure~\ref{fig:distdgl-gpu:low} and Figure~\ref{fig:distdgl-gpu:high} report the achieved speedups for all combinations of graph and graph partitioner under both configurations, respectively.} 
\blue{We observe that graph partitioning significantly speeds up GNN training. Specifically, a high quality partitioner such as \METIS{} leads to speedups of 1.09x, 1.48x, 1.37x, 1.69x, 1.47x, and 1.37x for the graphs \DI{}, \EN{}, \EU{}, \HO{}, \OR{}, and \PA{}, respectively in the low communication setting (see Figure~\ref{fig:distdgl-gpu:low}).} 
\blue{For the high communication setting, \METIS{} leads to higher speedups of 1.39x, 1.88x, 1.53x, 2.21x, 1.78x, and 1.56x for the same graphs, respectively (see Figure~\ref{fig:distdgl-gpu:high}). 
In some cases, even higher speedups were observed when using~\KAHIP{}.}

\blue{We conclude that \textbf{graph partitioning is an effective optimization in reducing GNN training time on GPUs.}}

%% file: sections/lessons_learnt.tex
\section{Lessons learned}
\label{sec:lessons}
In the following, we summarize our main findings and relate them to the research questions introduced in Section~\ref{sec:methodology}. 

\textbf{(1) Graph partitioning is effective in speeding up GNN training (RQ-1).} 
\blue{High-quality graph partitioning significantly enhances GNN training speeds. 
For \DISTGNN{}, we observe maximum speedups of 10.41x and average speedups of 4.36x when applying \HEPPP{}.
For \DISTDGL{}, we observe maximum speedups of 3.47x and average speedups of 1.37x when applying \KAHIP{}.
The speedups achieved for \DISTDGL{} are comparable to the speedups seen in distributed graph processing~\cite{hep,ease,twops}, whereas the speedups for \DISTGNN{} are higher. 
However, compared to \DISTGNN{}, \DISTDGL{} is a more mature and highly optimized system that is still under active development, indicating that there is less room for improvement with partitioning as for a less optimized system such as \DISTGNN{}.}  
Still, the results show the importance of partitioning for efficient distributed GNN training. 


\textbf{(2) Graph partitioning is effective in reducing the memory footprint (RQ-1).}
Graph partitioning effectively reduces memory demands.
We found a direct correlation between the replication factor and the memory requirement. 
Unlike in classical distributed graph processing where vertex states are small, GNNs exhibit large vertex states, often comprising large feature vectors and intermediate representations. 
Reducing the replication factor notably decreases memory consumption.
\blue{In many cases, advanced partitioning algorithms leading to smaller replication factors reduced the memory requirements by up to 85.1\% and on average by 51.4\%.}
Hence, the replication factor is pivotal for training a GNN \mbox{within a limited memory budget.}

\textbf{(3) Although classical partitioning metrics are relevant for predicting the performance of GNN training, further aspects should be considered (RQ-2).} 
The replication factor in DistGNN strongly correlates with network communication and memory overhead. 
In scenarios where different partitioners yield similar replication factors, vertex balancing emerges as crucial. 
Vertex balance aligns well with memory utilization balance, vital for memory-intensive GNN training, providing a key observation since most edge partitioners prioritize balancing edges and do not focus on vertex balancing.
This motivates including vertex balance into the edge partitioning problem definition. 
 
\textbf{(4) GNN parameters influence the effectiveness of graph partitioning (RQ-3).}
Unlike in traditional distributed graph processing, GNN training incorporates parameters such as number of layers, number of hidden dimensions, and batch size, alongside attached graph features. 
For DistDGL, we observed that GNN parameters influence the \blue{effectiveness of graph partitioning}.
Graph partitioning is effective, especially if the feature vectors are large and the hidden dimensions are low. 
We also found that if the feature size is large and the mini-batch size increases, the effectiveness markedly increases. 
For DistGNN, while training runtime is less influenced by these parameters, memory overhead efficiency improves with increases in feature size, hidden dimension, or number of layers.

\textbf{(5) The scale-out factor influences the effectiveness of graph partitioning (RQ-4).}
For DistDGL, we observed that in most cases, the effectiveness of graph partitioning slightly decreases when scaling out to more machines, while for DistGNN the effectiveness increases.
We found that for vertex partitioning (DistDGL) the cut-size of the partitioners increases more sharply than random partitioning.  
This results in a smaller effectiveness in terms of memory, communication and speedup. 
This is different from edge partitioning (DistGNN) where we observed an opposite trend: With more machines, the cut-size of the partitioners increases less sharply than for random partitioning and the effectiveness increases. 

\textbf{(6) Graph partitioning can usually be amortized by faster GNN training (RQ-5).}
Our findings for both DistDGL and DistGNN suggest that the invested graph partitioning time can in many cases be amortized already after a few epochs, making graph partitioning an important optimization for distributed GNN training.

\textbf{(7) Graph partitioning and sampling are interdependent.}
Our observations indicate that edge-cut and partitioning balance metrics do not always adequately represent computational balance and communication costs, since both can be influenced by sampling. 
These findings suggest a potential avenue for research on customizing sampling based on partitioning and vice versa.\vspace{-0.65mm}

%% file: sections/related-work.tex
\section{Related Work}
\label{sec:related-work}
Different studies \cite{survey.1.vldb.2017,survey.2.vldb.2018,survey.3.vldb.2018,survey.4.sigmod.2019} investigated how graph partitioning influences the performance of distributed graph processing. 
\citeauthor{survey.1.vldb.2017}~\cite{survey.1.vldb.2017} study graph partitioners available in GraphX~\cite{graphx}, PowerGraph~\cite{powergraph}, and PowerLyra~\cite{powerlyra} for graph analytics. 
\citeauthor{survey.2.vldb.2018}~\cite{survey.2.vldb.2018} study streaming graph partitioners and compare them in a graph processing framework based on Apache Flink~\cite{flink} for graph analytics.   
\citeauthor{survey.3.vldb.2018}~\cite{survey.3.vldb.2018} investigate the influence of different partitioning strategies in D-Galois for graph analytics workloads.
\citeauthor{survey.4.sigmod.2019}~\cite{survey.4.sigmod.2019} study streaming graph partitioners for graph analytics with PowerLyra and graph query workloads with JanusGraph~\cite{janus}.
The studies focus only classical graph workloads. 
However, distributed GNN training is different. 
First, GNN training leads to large memory and communication overheads. 
Huge feature vectors and large intermediate states are computed, stored, and sent over the network. 
Second, the computations consist of computationally expensive neural network operations. 
Third, GNN workloads are characterized by GNN parameters such as the number of layers and hidden dimensions.
Forth, mini-batch-based training has a complex data loading phase consisting of distributed multi-hop sampling followed by a communication intensive feature loading phase.


Graph partitioning is a vibrant research area and many different approaches exist~\cite{sandersschulz2013,adwise,twops,hep,dbh,hdrf,ne,dne,sheep,metis,windowvertexpartitioning,restreaming,schulz,ldg,spinner,xtrapulp,scotch,gcnsplit,cusp,ease}. 
See~\cite{schulzsurvey} for a recent survey about graph partitioning. 
We selected representative state-of-the-art streaming and in-memory graph partitioners for edge and vertex partitioning.

Many distributed graph neural network systems exist~\cite{DistGNN,distdgl,dorylus,bytegnn,su2021adaptive,deepgalois,p3}. A recent survey~\cite{surveyjana} gives an overview of different systems. 
We extend this research by experimentally investigating the effectiveness of graph partitioning for distributed GNN training.\vspace{-0.65mm}

%% file: sections/conclusion.tex
\section{Conclusions}
\label{sec:conclusion}
We performed an experimental evaluation to investigate the effectiveness of graph partitioning for distributed GNN training.
We showed that graph partitioning is an essential optimization for distributed GNN training and that different factors such as GNN parameters, the scale-out factor, and the feature size can influence the effectiveness of graph partitioning for GNN training, both in terms of memory footprint and training time. 
Further, we found that invested partitioning time can be amortized by reduced GNN training time.
Based on our findings, we conclude that graph partitioning has great potential to make GNN training more effective. 
We hope our research can spawn the development of even more effective graph partitioning algorithms in the future.  